\documentclass[aps,prd,pdflatex,reprint,showpacs,superscriptaddress]{revtex4-2}
\usepackage{amsmath}
\usepackage{amssymb}
\usepackage{amsthm}
\usepackage{bbm}
\usepackage{graphicx}
\usepackage{hyperref}
\usepackage{physics}
\usepackage{systeme}
\usepackage{times}
\usepackage{epstopdf}
\usepackage{float}
\usepackage{xcolor}
\usepackage[normalem]{ulem}

\hypersetup{
  colorlinks=true,linkcolor=blue,citecolor=blue,
  filecolor=blue,urlcolor=blue,breaklinks=true
}

\newcommand{\orcid}[1]{\href{https://orcid.org/#1}
{\includegraphics[width=7pt]{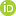}}}

\begin{document}
\title{Effects of rotation and Coulomb type potential on the spin-1/2 Aharonov-Bohm problem}
\author{M\'{a}rcio M. Cunha\orcid{0000-0001-9244-4669}}
\email{marciomc05@gmail.com}
\affiliation{
        Departamento de F\'{i}sica,
        Universidade Federal do Maranh\~{a}o,
        65085-580 S\~{a}o Lu\'{i}s, Maranh\~{a}o, Brazil
      }
\author{Fabiano M. Andrade\orcid{0000-0001-5383-6168}}
\email{fmandrade@uepg.br}
\affiliation{
  Departamento de Matem\'{a}tica e Estat\'{i}stica,
  Universidade Estadual de Ponta Grossa,
  84030-900 Ponta Grossa, Paran\'{a}, Brazil
}
\affiliation{
  Programa de P\'os-Gradua\c{c}\~{a}o Ci\^{e}ncias/F\'{i}sica,
  Universidade Estadual de Ponta Grossa,
  84030-900 Ponta Grossa, Paran\'a, Brazil
}

\author{Edilberto O. Silva\orcid{0000-0002-0297-5747}}
\email{edilberto.silva@ufma.br}
\affiliation{
Departamento de F\'{i}sica,
Universidade Federal do Maranh\~{a}o,
65085-580 S\~{a}o Lu\'{i}s, Maranh\~{a}o, Brazil
}
\date{\today }

\begin{abstract}
In this work, we investigate how both rotation and a Coulomb potential affect the quantum mechanical description of a spin-$1/2$ particle in the presence of the Aharonov-Bohm effect. We employ the method of the self-adjoint extensions in the framework of the Pauli-Schr\"{o}dinger equation. We discuss the role of the spin degree of freedom on this problem, find the energy spectrum, and investigate the results in detail.
\end{abstract}

\maketitle

\section{Introduction}
\label{sec:intro}

The Coulomb interaction has a fundamental relevance in describing
several physical systems, covering elementary aspects regarding the
electromagnetic theory.
Also, it is relevant to study systems that are in the scope of atomic
and nuclear physics.
In non-relativistic quantum mechanics, the Coulomb potential constitutes
a basic example in which we can solve the Schr\"{o}dinger equation
explicitly, in a complete and closed form \cite{Book.Muller.IQM.2012}.
More precisely, the Coulomb potential describes the interaction between
the electron and the nucleus for hydrogen-like atoms.

Due to its relevance, studying the quantum mechanical description of
other systems of interest in the presence of the Coulomb interaction
still is an attractive issue nowadays.
Besides the usual methods, it is possible, for example, to obtain exact
solutions for the one-dimensional Schr\"{o}dinger equation through the
Laplace transform \cite{JPA.2000.33.9265}.
The nonlinear logarithmic Schr\"{o}dinger equation in the presence of
the Coulomb potential also constitutes an example of a solvable problem
in this context \cite{JPC.2018.2.075014}.
Regarding its application, Coulomb-type potentials can model a wide
range of systems in condensed matter physics, like finite crystals
\cite{JPCM.1999.11.6159} and many-body Hamiltonians of
nanoscale devices \cite{PRB.2009.79.075315}, for example.
Recent contributions in the literature investigate the role of the
Coulomb interaction in the quantum mechanical description of a given
system.
In particular,  we can cite the study of the quantum Drude model
\cite{PRB.2016.94.115106} and the investigation of decoherence
mechanisms which are mediated by the Coulomb interaction
\cite{NJP.2020.22.063039}.

Concerning the fundamental aspects of quantum mechanics, another essential issue that attracts interest in several research lines consists of examining the importance of the Aharonov-Bohm effect (hereafter, AB effect) in the quantum dynamics of a system.
The seminal work of Aharonov and Bohm provided a new understanding of the significance of the vector potential in the quantum domain \cite{PR.1959.115.485}.
As a historical and pedagogic remark, it is worth mentioning that Aharonov and Bohm have proposed both a magnetic and an electric version of that effect.
In the present work, we are interested in the case of the magnetic one.
Also, let us have in mind that the AB effect can emerge in two different situations: in the context of interference of the wave function of electrons, for example, when a magnetic flux is present, and also in the case of electrons which are constrained to move in a limited region of space.
 We can associate the appearance of geometrical phases to the first case, while in the second one, the AB effect alters the bound state energies of the system \cite{PRA.1981.23.360}.
 These two situations can occur in a wide variety of physical systems.
 For instance, in condensed matter systems like graphene and carbon nanotubes, the AB effect introduces oscillations in the energy gap of these structures \cite{PRB.2010.81.195441,RSCAdv.2015.5.45551}.
Also, it is possible to create AB interferometers to perform transport measurements in graphene \cite{JPCM.2018.30.485302}.
 The AB effect also can occur in the relativistic domain \cite{PLA.1989.140.105}.
It is relevant to mention that the concepts behind the AB effect have inspired the study of analogs such as a gravitational one \cite{PRL.2012.108.230404} and a version for neutral particles named the Aharonov-Casher effect \cite{PLA.1991.154.93}. Beyond these topics, we can dedicate attention to aspects concerning
fundamental issues involving the AB effect.
Currently,  there still exists a debate on the origin, interpretations
and  implications of the AB effect
\cite{PRA.2016.93.042110,PRA.2017.95.052123,Entropy.2018.20.465,FP.2018.48.837}. There are also relevant works in the literature on the AB effect applied to other physical systems, for example, problems involving spin and pseudo-spin symmetries \cite{EPJC.2019.79.596,CTP.2015.64.637,CPC.2014.38.013101,PLA.2010.374.4303,FBS.2010.48.171,CTP.2012.58.807,MPLA.2010.25.2447,IJP.2014.88.405,EPJC.2015.75.321}, topological defects \cite{IJMPD.2018.27.1850005,GRG.2019.51.120,AdP.2010.522.447,PRD.2020.102.105020,AoP.2015.362.739,PRD.2012.85.041701,AdHEP.2021.2021.6709140,PRD.2012.85.041701}, thermodynamic aspects \cite{EPJP.2019.134.495,FPS.2015.56.115,EPJP.2020.135.691,EPJD.2020.74.159,JLT.2021.202.83,PE.2021.131.114710,PLJP.2021.136.432,IJT.2021.42.138}, $\kappa$-deformed algebra \cite{PLB.2013.719.467,PLB.1995.359.339}, Lorentz symmetry violation \cite{JPG.2012.39.55004,EPJP.2020.135.656,PRD.2011.83.125025,AdP.2014.526.514} and Duffin-Kemmer-Petiau (DKP) formalism \cite{PPNL.2019.16.195,AdHEP.2018.2018.1031763,JPA.51.2017.035201,MPLA.2021.36.2150059}.
Another topic of central importance in this context refers to the
description of the AB effect by taking into account the electron spin
degree of freedom \cite{PRL.1990.64.503}.
As we shall see in more detail below, this problem demands particular
mathematical tools for adequate treatment \cite{FP.2019.7.00175}.
 When a magnetic field is present, we know that the spin degree of freedom is responsible for lifting the degeneracy of the energy levels of a particle due to the Zeeman interaction.
 A similar effect occurs when a system is rotating.
More specifically, there is a coupling between the spin degree of freedom and the rotation known as spin-rotation coupling.
That coupling also breaks the degeneracy between the different spin eigenvalues.
In the context of interferometry, we can consider it as
 a quantum mechanical extension of the Sagnac effect \cite{NPJQI.2020.6.1}.
In solid state physics, spin-rotation coupling can be a route for the generation of spin currents \cite{SSC.2014.198.52}.
 Another interesting situation occurs in the framework of linearized general relativity, consisting of a spin-rotation-gravity coupling.
That quantity allows defining a gravitomagnetic Stern-Gerlach force \cite{Entropy.2021.23.445}.

In general, studying the effects of rotation on quantum systems is not restricted to spin-rotation coupling, and others aspects are also relevant.
Another contribution from rotation appears on the orbital angular momentum, which also appears in the classical scenario.
 From gravitation and cosmology until the condensed matter physics, rotation effects constitute a fruitful issue for investigation.
 In this context, rotation can be inherent to the system.
It occurs in the case of neutron stars, black holes, and fullerenes, for instance.
In a neutral star, rotation powers its electromagnetic activity \cite{Universe.2020.6.15}.
Similar behavior occurs in the case of black holes, in which rotation creates processes of ionization and irradiation \cite{Universe.2020.6.26}.
Fullerenes spin rapidly at room temperatures, and a laser pulse can transfer angular momentum to $C_{60}$, affecting lattice properties like phonons \cite{PRB.2021.104.L100302}.
Such molecules also are present in fullerite crystals, having a central role in the crystal behavior \cite{EPJP.2021.136.1}.
In rotating semiconductors, impressive features emerge.
In particular, these systems present a purely quantum mechanical effect, which consists of breaking the equivalence of clockwise/counterclockwise rotations, modifying their mechanical and transport properties \cite{Symmetry.2021.13.1569}.
Another example of a system in which this clockwise/counterclockwise asymmetry arises refers to the AB effect for excitons in a semiconductor quantum ring \cite{PRB.2015.91.235308}.

Considering that both AB effect and rotation can reveal genuinely quantum mechanical effects in a system, it is interesting to investigate the dynamics of a quantum system in the presence of these two features.
Besides, due to the role of the spin degree of freedom in producing a Zeeman interaction and the spin-rotation coupling, it is relevant to include it in the problem.
Then, in this work, we study the spin-$1/2$ Aharonov-Bohm problem
subjected to rotation and a Coulomb-type interaction.
The introduction of the Coulomb potential is also motivated by its vital importance in modeling quantum systems.
Since we are interested in examining the nonrelativistic scenario, we work with the Pauli-Schr\"{o}dinger equation to accommodate the presence of the spin degree of freedom.

The organization of the paper is as follows.
In Sec. \ref{sec:Pauli}, we introduce the Pauli-Schr\"{o}dinger equation for a rotating frame in the presence of the AB effect and a Coulomb-type potential.
By employing the method of the self-adjoint extensions, we solve the corresponding differential equation, obtaining the energy spectrum for the problem.
Then, in Sec. \ref{sec:discussion}, we focus our attention on analyzing the results, through several graphs of the energy spectrum as a function of the physical parameters of the system.
We make our conclusions in Sec. \ref{sec:conclusions}.

\section{The Pauli-Schr\"{o}dinger equation in a rotating frame}
\label{sec:Pauli}

The equation that describes the motion of a nonrelativistic spin-1/2
particle in the presence of an electromagnetic field in a rotating frame
is the Pauli-Schr\"{o}dinger equation with the inclusion of the Zeeman
energy term, the term describing inertial effects plus the spin-rotation
coupling \cite{PhysRevB.84.104410}.
The corresponding equation is
\begin{equation}
  i\hslash \frac{\partial \psi }{\partial t}=
  \left[
    \frac{\boldsymbol{\pi }^{2}}{2m_{e}}+V\left( r\right)
    -\boldsymbol{\mu}\cdot \mathbf{B}
    -\mathbf{\Omega} \cdot \left(\mathbf{r}\times\boldsymbol{\pi}
      +\mathbf{S}\right)
  \right]\psi,
  \label{sp}
\end{equation}
where $\boldsymbol{\mu}=\left(e\hbar \mathbf{S}/m_{e}c\right)$ is
the magnetic moment, $m_{e}$ is the electron mass, $c$ is the speed of
light, $e$ is the electric charge, $\hslash $ is the Planck constant, $
\mathbf{\Omega}$ is the rotation frequency with respect to an inertial
frame, $\mathbf{A}$ is the vector potential, $V\left(r\right)$ is the
scalar potential and $\boldsymbol{\pi}=\mathbf{p}-e\mathbf{A}/c$ is the mechanical momentum.
The spin operator $\mathbf{S}$ is $\hbar \boldsymbol{\sigma}/2$.
Also, we assume that the rotation velocity $
\mathbf{\Omega \times r}$ is uniform, where $\mathbf{r}$ is the position
vector from the origin at the rotation axis.
Note that the explicit
dependence relative to the spin degree of freedom in Eq.
(\ref{sp}) is through the magnetic interaction $\boldsymbol{\mu }\cdot \mathbf{B}$ and the
spin-rotation coupling $\mathbf{\Omega }\cdot \mathbf{S}$.
 For the field configuration, we consider the magnetic field of an infinitely long, infinitesimally thin
solenoid along the z-axis
\begin{equation}
\;e\mathbf{B}=e\left( \mathbf{\nabla }\times \mathbf{A}\right) =\left(
0,0,-\phi \frac{\delta (r)}{r}\right) ,  \label{b}
\end{equation}
with $r$ being the two-dimensional radius vector.
In the Coulomb gauge, the
magnetic field $\mathbf{B}$ is due to the potential vector
\begin{equation}
e\mathbf{A}=\left( 0,-\frac{\phi }{r},0\right) ,  \label{a}
\end{equation}
where $\phi =\Phi /\Phi _{0}$ is related to the Aharonov-Bohm flux along the
solenoid, in which $\Phi $ denotes the magnetic flux and $\Phi _{0}=2\pi /e$
indicates the quantum of magnetic flux.
The scalar potential $V(r)$ in Eq. (\ref{sp}) is an attractive Coulomb-type potential given by
\begin{equation}
V(r) =-\frac{\eta}{r},  \label{pc}
\end{equation}
where $\eta$ is a real positive parameter describing the strength of the potential.
When incorporated into the present model, the potential (\ref{pc}) should be interpreted as being cylindrically symmetric, and it also allows us to obtain analytical solutions to the Pauli-Schrödinger equation (\ref{sp}).
For the rotation frequency, we specialize to the case $\mathbf{\Omega}=\left(0,0,\Omega \right)$, which is related to the rotation
velocity as $\mathbf{\Omega }\times \mathbf{r}=\Omega r\,\boldsymbol{\hat{\varphi}}$.
Before solving Eq.
(\ref{sp}), we can note that the wave function $\psi$ is an eigenfunction of $\sigma ^{z}$, whose eigenvalues are known to be $s=\pm 1$, satisfying $\sigma ^{z}\psi =s\psi =\pm \psi $.
For a stationary solution of
energy $\mathcal{E}$ of the form
\begin{equation}
\psi \left( r,\varphi \right) =e^{-\frac{i\mathcal{E}}{\hbar }}e^{im\varphi }
\mathcal{F}(r),  \label{fo}
\end{equation}
where $m=0,\pm 1,\pm 2,\pm 3,\ldots $ is the angular momentum quantum
number, we obtain the radial equation of motion
\begin{equation}
\mathcal{H}\mathcal{F}_{m}(r)=\kappa^{2}\mathcal{F}_{m}(r),  \label{re}
\end{equation}
where
\begin{equation}
\mathcal{H}=\mathcal{H}_{0}+s\phi \,\frac{\delta \left( r\right)
}{r}  \label{oh}
\end{equation}
and
\begin{equation}
\mathcal{H}_{0}=-\frac{1}{r}\frac{d}{dr}\left( r\frac{d}{dr}\right) +\frac{
j^{2}}{r^{2}}-\frac{2m_{e}\eta
^{\prime}}{r},  \label{h0}
\end{equation}
\begin{align}
\kappa^{2}&=\frac{2m_{e}\mathcal{E}}{\hbar ^{2}}+\frac{2m_{e}\Omega }{\hbar}
\left( j+\frac{s}{2}\right),\label{k2}\\ j&=m+\phi, \;\; \eta^{\prime}=\frac{\eta}{\hbar^{2}}.
\label{j}
\end{align}
The radial equation (\ref{re}) shows up in several articles in the
literature and the operator $\mathcal{H}$ is not essentially
self-adjoint for all values of $j$.
Here, we characterize the family of self-adjoint extensions of
$\mathcal{H}_{0}$, replacing the $\delta$ function by a boundary condition
at the origin as established in Ref. \cite{JMP.26.2520.1985} (see also Refs. \cite{FP.2019.7.00175}).
This boundary condition is a mathematical limit that allows divergent
solutions for the Hamiltonian $\mathcal{H}_0$ at isolated regions,
provided they remain square-integrable.
For the present model, the access to such isolated regions is
conditioned by the relation \cite{PRD.50.7715.1994}
\begin{equation}
    | j| < \frac{1}{2},\label{rj}
\end{equation}
which evidences the region where the operator $\mathcal{H}_0$ is
essentially self-adjoint, i.e., $|j|\geq 1/2$.
From Eq. (\ref{j}), we see that the isolated regions depend on the values
of  $\phi$.
Then, by decomposing the magnetic quantum flux as
\begin{equation}
  \phi = N + \beta, \label{fq}
\end{equation}
with $N$ being the largest integer contained in $\phi$, and the quantity $\beta$ being defined in the range
\begin{equation}
  0 \leq \beta < 1,
\end{equation}
we find an expression for the angular momentum quantum number $m$
satisfying the relation (\ref{rj}), which is given by
\begin{equation}
  -\frac{1}{2}-(N+\beta)< m < \frac{1}{2}-(N+\beta),\label{int}
\end{equation}
and it shows the range of $m$ in which $\mathcal{H}_0$ is not
self-adjoint.
In view of the above information, it is evident that the radial equation
(\ref{re}) must be solved taking into account the relation (\ref{rj}).
In this way, all the  self-adjoint extensions
$\mathcal{H}_{0,\lambda_{m}}$ of $\mathcal{H}_{0}$ are characterized by
the boundary condition at the origin \cite{JMP.26.2520.1985}
\begin{equation}
  \lambda_{m} \mathcal{F}_{0} = \mathcal{F}_{1},
  \label{bc}
\end{equation}
with $-\infty < \lambda_{m} \leq \infty$, $-1/2 < j < 1/2$ and the
boundary values are given by
\begin{align}
  \mathcal{F}_{0} = {}
  & \lim_{r \to 0^{+}}r^{|j|}\mathcal{F}_{m}(r),  \nonumber \\
  \mathcal{F}_{1} = {}
  & \lim_{r \to 0^{+}}\frac{1}{r^{|j|}}
    \left[
    \mathcal{F}_{m}(r)-\mathcal{F}_{0}\frac{1}{r^{|j|}}\right].
\end{align}
It is important to emphasize that when $\lambda_{m}=\infty$, it means
that we  are dealing with the free Hamiltonian without the point
interaction.
In this case, the wave function is regular at the origin, and the
original AB problem \cite{PR.115.485.1959} is recovered.
For any other values of $\lambda_{m}$, i.e., $|\lambda_{m}|<\infty$, the Hamiltonian characterizes a singularity at $r=0$ region.
As a result, the boundary condition admits a $r^{-|j|}$ singularity in the wave functions at this point \cite{PRA.77.036101.2008}.
This is a well-known result in the literature.

It can be shown that Eq.
(\ref{fo}) is of the confluent hypergeometric equation type, whose general solution is given in terms of the Kummer function $M\left(a,b,z\right)$ and the confluent hypergeometric function of the second kind $U\left(a,b,z\right)$.
A convenient expression for dealing with this solution is achieved by making use of properties allowing us to write the confluent hypergeometric function of the second kind in terms of Kummer functions (or confluent hypergeometric function of the first kind $_{1}F_{1}\left(a,b,z\right)$).
In this way, we express the solution of Eq.
(\ref{fo}) in the form
\begin{align}
  \mathcal{F}_{m}(x)= {}
  &
    a_{m}\,x^{| j| }e{^{-\frac{x}{2}}}\,_{1}F_{1}{
    \left( a,\,b,\,x\right) } \notag \\
  &+b_{m}\,x^{-| j| }{}e{^{-\frac{x
}{2}}}~_{1}F_{1}\left( a^{\prime },b^{\prime },x\right), \label{sg}
\end{align}
with
\begin{align}
x = {} &2\kappa r, \\
a = {} &\frac{1}{2}+| j| -\,{\frac{m_{e}\eta ^{\prime }}{
\kappa }},\;\;b=1+2| j| , \\
{a}^{\prime } = {} &\frac{1}{2}-| j| -\,{\frac{m_{e}\eta
^{\prime }}{\kappa }},\;\;b^{\prime }=1-2| j|,
\end{align}
where $a_{m}$ and $b_{m}$ are, respectively, the coefficients of the regular and irregular solutions at the origin.
Note that in the case of bound state solutions, it must be ensured that the function $e^{-x/2}$ is convergent when $x=\infty$.
Thus, we write the quantity $\kappa^{2}$ in Eq.
(\ref{k2}) as
\begin{equation}
\kappa =\sqrt{-\left[ \frac{2m_{e}\mathcal{E}}{\hbar ^{2}}+\frac{%
2m_{e}\Omega }{\hbar }\left( j +\frac{s}{2}\right) \right]},
\label{k2r}
\end{equation}
with the requirement that
\begin{equation}
\frac{2m_{e}\mathcal{E}}{\hbar ^{2}}+\frac{2m_{e}\Omega }{\hbar }\left(
j +\frac{s}{2}\right) <0  \label{exi}
\end{equation}%
to guarantee that $\kappa $ is a real quantity.
With the condition
(\ref{exi}), the exponential function $e{^{-x/2}}$ tends to zero when $x\rightarrow \infty $.
On the other hand, if
\begin{equation}
\frac{2m_{e}\mathcal{E}}{\hbar ^{2}}+\frac{2m_{e}\Omega }{\hbar }\left(
j +\frac{s}{2}\right) >0,
\end{equation}%
the exponential $e{^{-x/2}}$ oscillates when $x\rightarrow \infty$.

Our main goal is to obtain an expression for bound state energies.
We make that through two stages.
In the first one, we must substitute the solution (\ref{sg}) into the boundary condition (\ref{bc}), which should provide a relation between the coefficients $a_{m}$ and $b_{m}$.
In the second stage, we require that the solution be normalizable at large $r$.
From this condition,  we can obtain another relation between the coefficients $a_{m}$ and $b_{m}$.
Combining the two relations involving such coefficients makes it possible to find a secular equation that provides the energies of the bound states.
The two procedures are performed as follows.
Since the boundary condition is applied at the $r=0$ region, we must use series expansions up to second order in the argument to the functions $e^{-x/2}$ and $_{1}F_{1}{\left( a,\,b,\,x\right)}$.
After combining the results, we get
\begin{align}
  e{^{-\frac{x}{2}}} _{1}F_{1}{\left( a,\,b,\,x\right)}\approx
  &
    \frac{\left(x^{2}-4x+8\right) }{16b\left( b+1\right) }\notag\\
  &
    \times\left[ \left( a^{2}+a\right)
x^{2}+2\left( ax+b\right) \left( b+1\right) \right].
\end{align}
With this result, the solution (\ref{sg}) takes the form
\begin{align}
  \mathcal{F}_{m}(x) = {}
  &
    a_{m}\frac{\left(x^{2}-4x+8\right)
    x^{|j|}}{16b\left(b+1\right)}\notag \\
  &
   \times\left[ \left(a^{2}+a\right) x^{2}+2\left(ax+b\right)
\left( b+1\right) \right]   \notag \\
& +b_{m}\frac{\left(x^{2}-4x+8\right) x^{-| j|}}{
16b\left(b+1\right)}\notag \\&\times\left[\left(a^{\prime 2}+a^{\prime }\right)
x^{2}+2\left(a^{\prime}x+b^{\prime}\right) \left(b^{\prime}+1\right)
\right].\label{sn}
\end{align}
Substituting the solution (\ref{sn}) into the boundary condition (\ref{bc}) and separating the terms involving the limits that require some criterion for their realization, we find
\begin{align}
&b_{m}\,\left( 2\kappa \right) ^{-| j|} =\lambda _{m}\notag \\&\times
\left[ a_{m}\left( 2\kappa \right) ^{| j| }+b_{m}\left(
2\kappa \right) ^{1-| j| }\left( \frac{a^{\prime }}{
b^{\prime }}-\frac{1}{2}\right) \lim_{r\rightarrow 0^{+}}{}r^{1-2|
j| }\right]   \notag \\
& +\lambda _{m}\Bigg[ b_{m}\frac{1}{2}\left( 2\kappa \right) ^{2-|
j| }{}\left( \frac{1}{4}-\frac{a^{\prime }}{b^{\prime }}+\frac{
a^{\prime }}{b^{\prime }{}^{2}+b^{\prime }}+\frac{a^{\prime }{}^{2}}{
b^{\prime }{}^{2}+b^{\prime }}\right) \notag \\&\times\lim_{r\rightarrow
0^{+}}r^{2-2| j| }\Bigg].\label{cn2}
\end{align}
It is possible to recover results from the literature through Eq.
(\ref{cn2}).
For this purpose, it is more convenient to write it in the form
\begin{align}
&b_{m}\left( 2\kappa \right) ^{-2| m+\phi | }
=a_{m}\lambda _{m}{}\notag\\&-\lambda _{m}b_{m}\left( 2\kappa \right) ^{-2|
m+\phi | }\left( \frac{2\eta ^{\prime }m_{e}}{2| m+\phi
| -1}\right) \lim_{r\rightarrow 0^{+}}{}r^{1-2| m+\phi
| }  \notag \\
& +b_{m}\frac{\lambda _{m}}{4}\left( \frac{\kappa ^{2}-2\kappa
^{2}| m+\phi | +4\eta ^{\prime }{}^{2}m_{e}^{2}}{
2| m+\phi | ^{2}-3| m+\phi | +1}
\right) \left( 2\kappa \right) ^{-2| m+\phi |
}{}\notag \\ &\times\lim_{r\rightarrow 0^{+}}r^{2-2| m+\phi | }.\label{sln2}
\end{align}

In the absence of a Coulomb potential, it is known that the wave functions are given in terms of Bessel functions.
Since it is possible to express the Bessel functions in terms of the confluent hypergeometric function, we can recover the expected result by setting $\eta^{\prime}=0$ in Eq.
(\ref{sln2}).
We obtain
\begin{align}
  a_{m}\lambda_{m}= {}
  &
    b_{m}\left(2\kappa \right)^{-2| m+\phi
    |}\notag\\
  &
    \times\left[ 1-\lambda_{m}\frac{\kappa ^{2}}{4\left( 1-|
m+\phi | \right)}{}\lim_{r\rightarrow 0^{+}}r^{2-2|
m+\phi |}\right].\label{ptc}
\end{align}
The occurrence of a regular solution directly implies that $b_{m}=0$ (or $\lambda_{m}=0$) in Eq.
(\ref{ptc}).
In contrast to this case, the occurrence of a singular solution demands that $\left \vert m+\phi \right \vert<1$, and the relation between the coefficients yields $a_{m}/b_{m}\sim \lim_{r\rightarrow 0^{+}}r^{2-2| m+\phi |}$.
As expected, this result coincides with Eq.
(15) of Ref.
\cite{PRD.1993.48.5935} by making $r=R$.
Furthermore, it is also important to mention that Eq.
(\ref{ptc}) reveals us that the range of $\left \vert m+\phi \right \vert$ is increased in the absence of the Coulomb potential.
Equation (\ref{ptc}) also can be compared with Eq.
(64) of Ref.
\cite{AoP.2013.339.510}.
For this to be accomplished, we must restore the parameters $\kappa \rightarrow ik$ and $m+\phi \rightarrow \left[m+\phi +\left(1-\alpha \right)/2\right]/\alpha$ and then multiplying the resulting equation by $\left(2\kappa \right)^{-|m+\phi|}$.
This leads to the expression
\begin{align}
  a_{m}\lambda_{m}= {}
  &
    b_{m}\left(2ik\right) ^{-\frac{2}{\alpha }
    \left[ m+\phi +\frac{1}{2}\left( 1-\alpha \right) \right] }\notag\\
  &
    \times\Bigg[ 1+\frac{
\lambda _{m}k^{2}}{4\left( 1-\frac{1}{\alpha }\left[m+\phi +\frac{1}{2}
    \left( 1-\alpha \right) \right] \right) }\notag \\
  &
    \times\lim_{r\rightarrow 0^{+}}r^{2-\frac{
2}{\alpha }\left[ m+\phi +\frac{1}{2}\left( 1-\alpha \right) \right] }\Bigg],
\end{align}
which is just Eq.
(64) of Ref.
\cite{AoP.2013.339.510}.

Returning to our problem and analyzing the limits in Eq.
(\ref{sln2}), we find that the relevant one has the lowest power in $r$, which allows us to neglect $\lim_{r\rightarrow
0^{+}}r^{2-2| j|}$.
After simplifying the resulting expression, we obtain
\begin{equation}
a_{m}\lambda_{m}\left( 2\kappa \right) ^{2| j| }=b_{m}
\left[ 1\,+\left( \frac{2\lambda _{m}m_{e}\eta ^{\prime }}{1-2|
j| }\right) \lim_{r\rightarrow 0^{+}}{}r^{1-2|
j| }\right].
 \label{ns}
\end{equation}
Notice that $\lim_{r\rightarrow 0^{+}}{}r^{1-2|j|}$ is convergent if
$|j|<1/2$ and diverges if $|j|\geq 1/2$, thus revealing the condition
for the occurrence of a singular solution at the origin.
In other words, this means that for $|j|\geq 1/2$ the coefficient
$b_{m}$ is zero, which implies that only the regular solution at the
origin should be considered in the solution
\cite{AoP.2013.339.510,JPG.2013.40.075007,TMP.2013.175.637,
EPJC.2013.73.2548,AoP.2008.323.1280,TMP.2009.161.1503,EPJC.2014.74.2708}.
It is also known that the occurrence of singular solutions is a
consequence of the fact that the operator $\mathcal{H}_0$ is not
self-adjoint for the condition $|j|<1/2$.

Let us now perform the second stage in order to find the second relation between $a_{m}$ and $b_{m}$.
Since the requirement to ensure that the wave function is normalizable is $\lim_{x\rightarrow \infty }\mathcal{F}_{m}(x)=0$, we use the asymptotic representation of the function $_{1}F_{1}{\left( a,\,b,\,x\right)}$ for large $x$

\begin{equation}
_{1}F_{1}{\left( a,\,b,\,x\right) }=\frac{\Gamma (b)}{\Gamma (a)}
e^{x}x^{a-b}+\frac{\Gamma (b)}{\Gamma (b-a)}(-x)^{-a},  \label{f1a}
\end{equation}
which substituted into solution (\ref{sg}), results
\begin{multline}
  \lim_{x\rightarrow \infty }\mathcal{F}_{m}(x) = \\
  a_{m}\,x^{| j| }e{^{-\frac{x}{2}}}\,
  \left[\frac{\Gamma (b)}{\Gamma (a)}e^{x}x^{a-b}
    +\frac{\Gamma (b)}{\Gamma (b-a)}(-x)^{-a}\right]   \notag \\
 +b_{m}\,x^{-| j| }{}e{^{-\frac{x}{2}}}~\left[ \frac{
\Gamma (b^{\prime })}{\Gamma (a^{\prime })}e^{x}x^{a^{\prime }-b^{\prime }}+
\frac{\Gamma (b^{\prime })}{\Gamma (b^{\prime }-a^{\prime })}
(-x)^{-a^{\prime }}\right] \notag \\ =0.
\end{multline}
\begin{figure*}[t]
  \centering
  \includegraphics[scale=0.22]{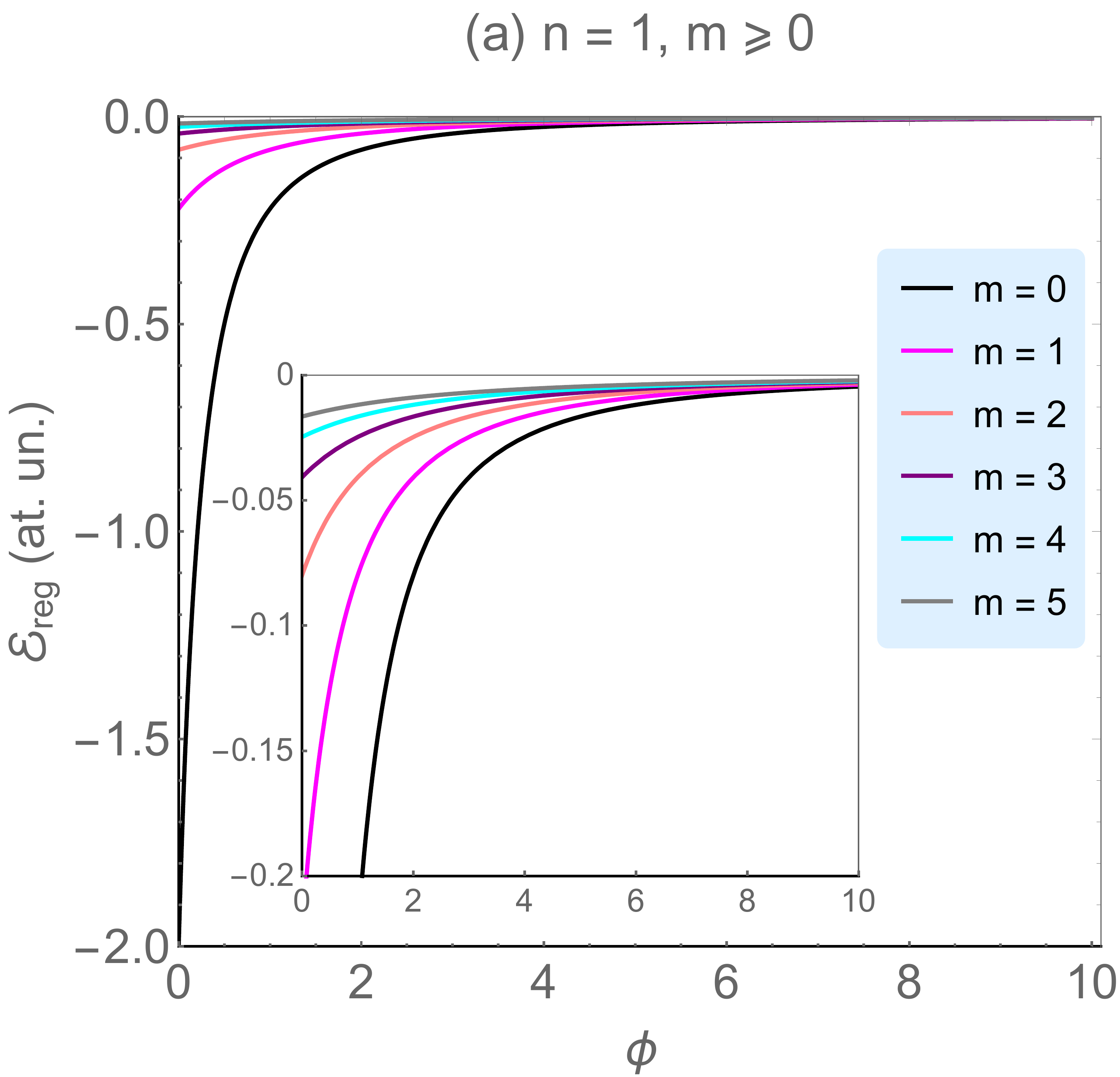}\qquad
  \includegraphics[scale=0.31]{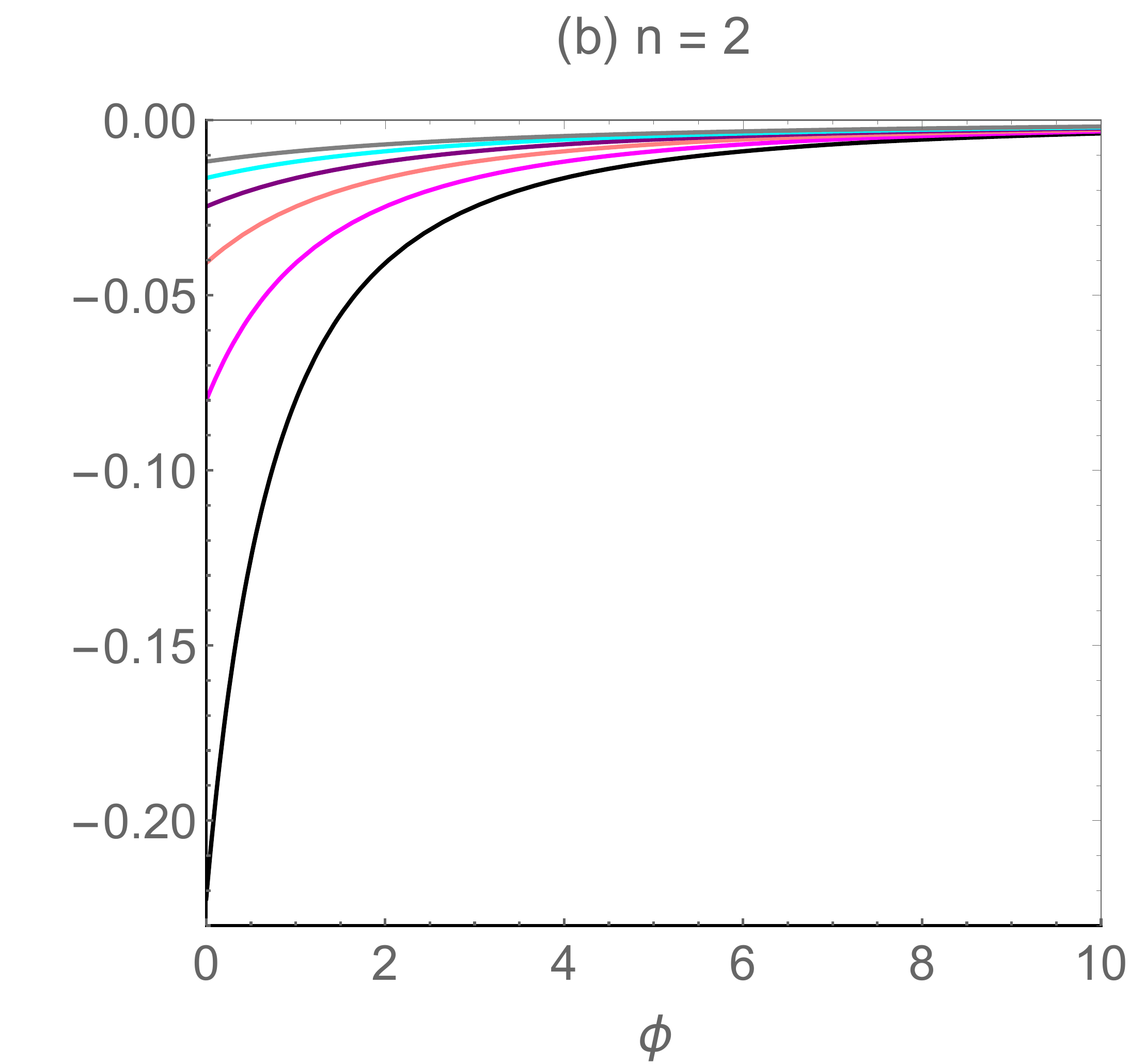}
  \caption{(Color online)
    Energy of the regular case (Eq.(\ref{Ereg})) as a function of $\phi$
    without taking into account rotation effects ($\Omega=0$) for
    $m\geqslant0$.
    In panel (a), we plot for $n=1$ and in panel (b) for $n=2$.
    The energy of the ground state for a specific value of $m$ tends to
    zero as the flux increases.
    For $n=2$, we see clearly that the effects manifested in (a) tend to
    occur more rapidly.}
  \label{fig:fig1}
\end{figure*}
Neglecting the terms involving functions that diverge on the $x\rightarrow \infty$ limit and simplifying the result, we arrive at the following relation:
\begin{equation}
a_{m}\Gamma (b)\Gamma (b^{\prime }-a^{\prime })+b_{m}\Gamma (b^{\prime
})\Gamma (b-a)=0,
\end{equation}
which can be written more explicitly as
\begin{multline}
  a_{m}\Gamma \left( 1+2| j| \right)
  \Gamma \left( \frac{1}{2}-l_{-}\right)\\
  +b_{m}\Gamma \left( 1-2| j| \right)
  \Gamma \left( \frac{1}{2}+l_{+}\right) = 0, \label{rl2}
\end{multline}
where we have defined the parameters
$l_{+}=| j| +\,{m_{e}\eta ^{\prime }/\kappa }$ and
$l_{-}=| j| -\,{m_{e}\eta ^{\prime }/\kappa }$.
Equation (\ref{rl2}) is the second relation between $a_{m}$ and $b_{m}$
that we need.
\begin{figure*}[!t!]
  \centering
  \includegraphics[scale=0.21]{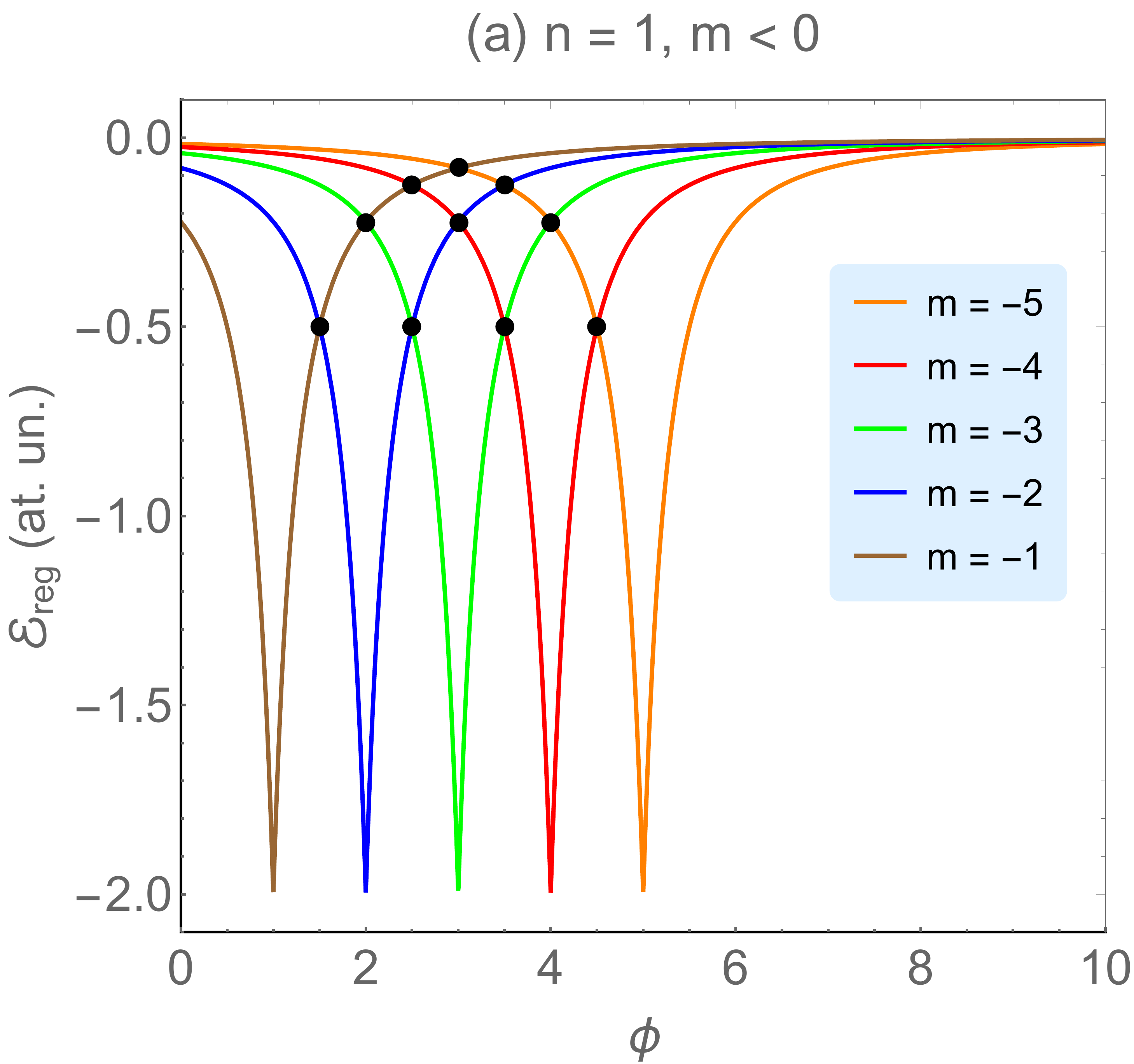}\vspace{1cm}
  \includegraphics[scale=0.29]{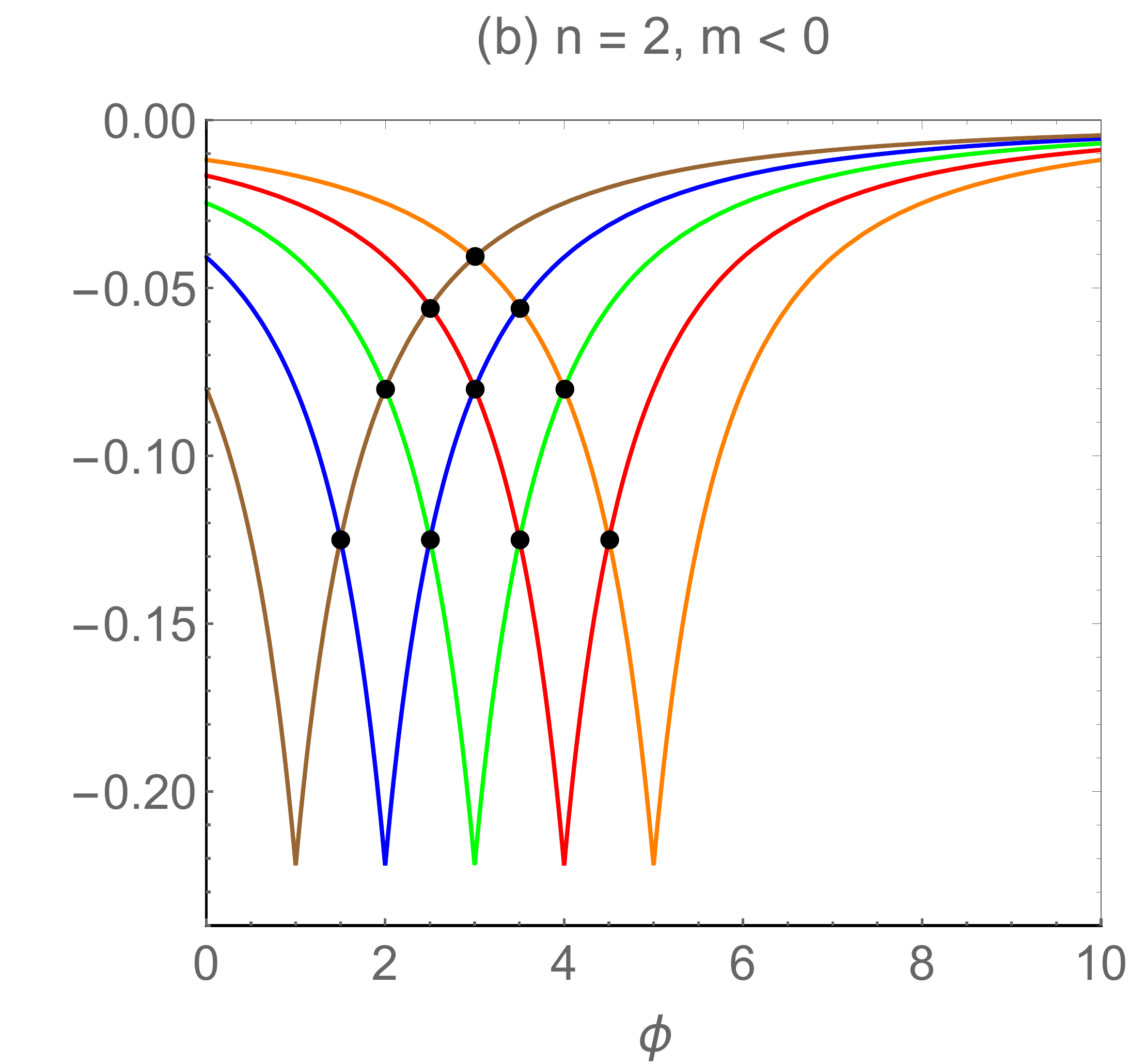}
  \caption{(Color online)
    Energy of the regular case (Eq. (\ref{Ereg})) as a function of the
    Aharonov-Bohm flux without taking into account rotation effects
    ($\Omega=0$) for $m<0$.
    In panel (a), we plot for $n=1$ and in panel (b) for $n=2$.
    In (a), $\mathcal{E}_{\rm reg}$ increases with the flux and has
    maximum magnitude for integer values of $\phi$.
    The maximum value of $\mathcal{E}_{\rm reg}$ is the same for all
    states with $m<0$.
    In the first excited state, we verify that only the location of the
    degenerate states is changed on the energy scale (panel (b)).
  }
  \label{fig:fig2}
\end{figure*}
\begin{figure*}[!t!]
  \centering
  \includegraphics[scale=0.3]{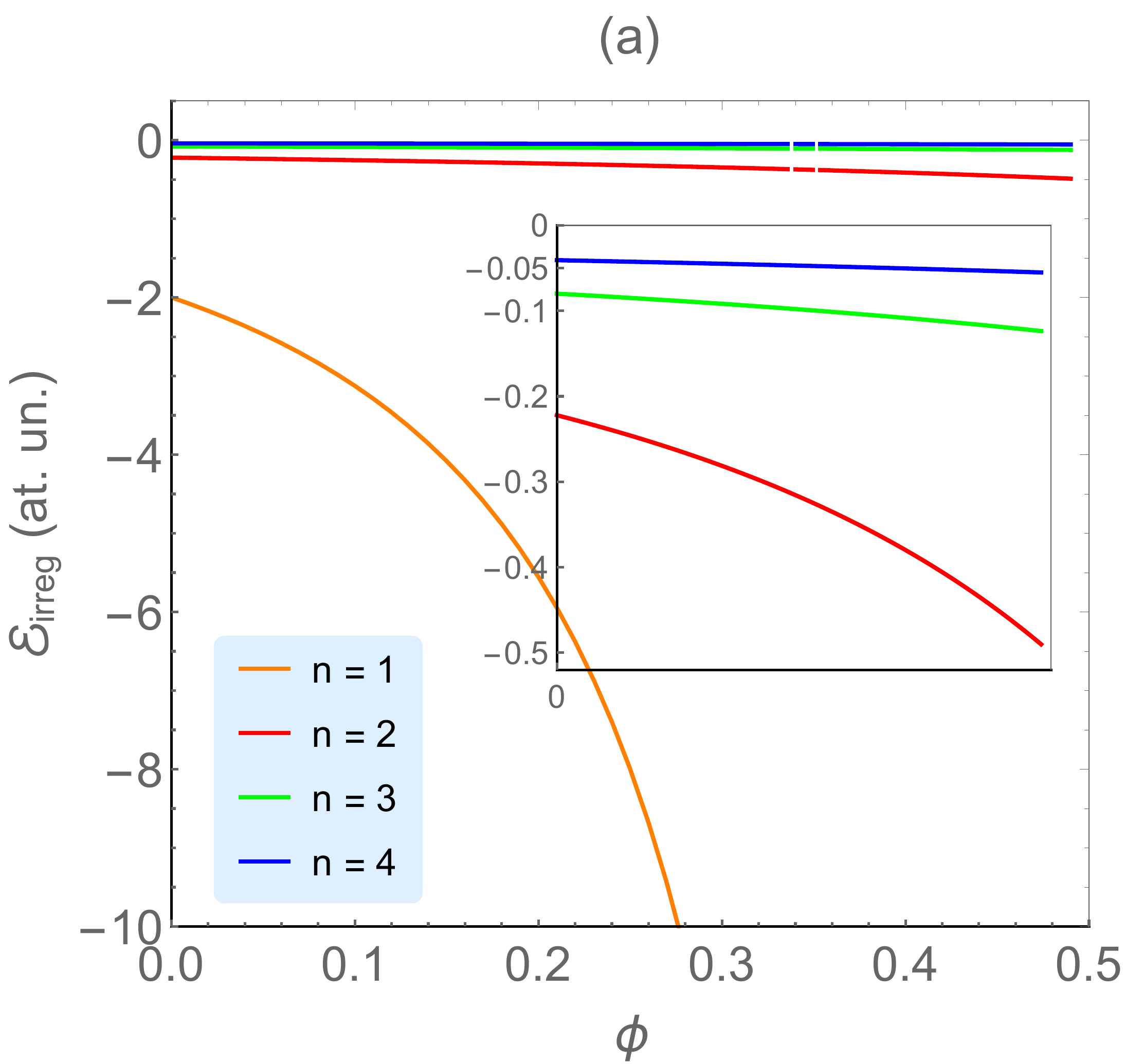}\vspace{1cm}
  \includegraphics[scale=0.32]{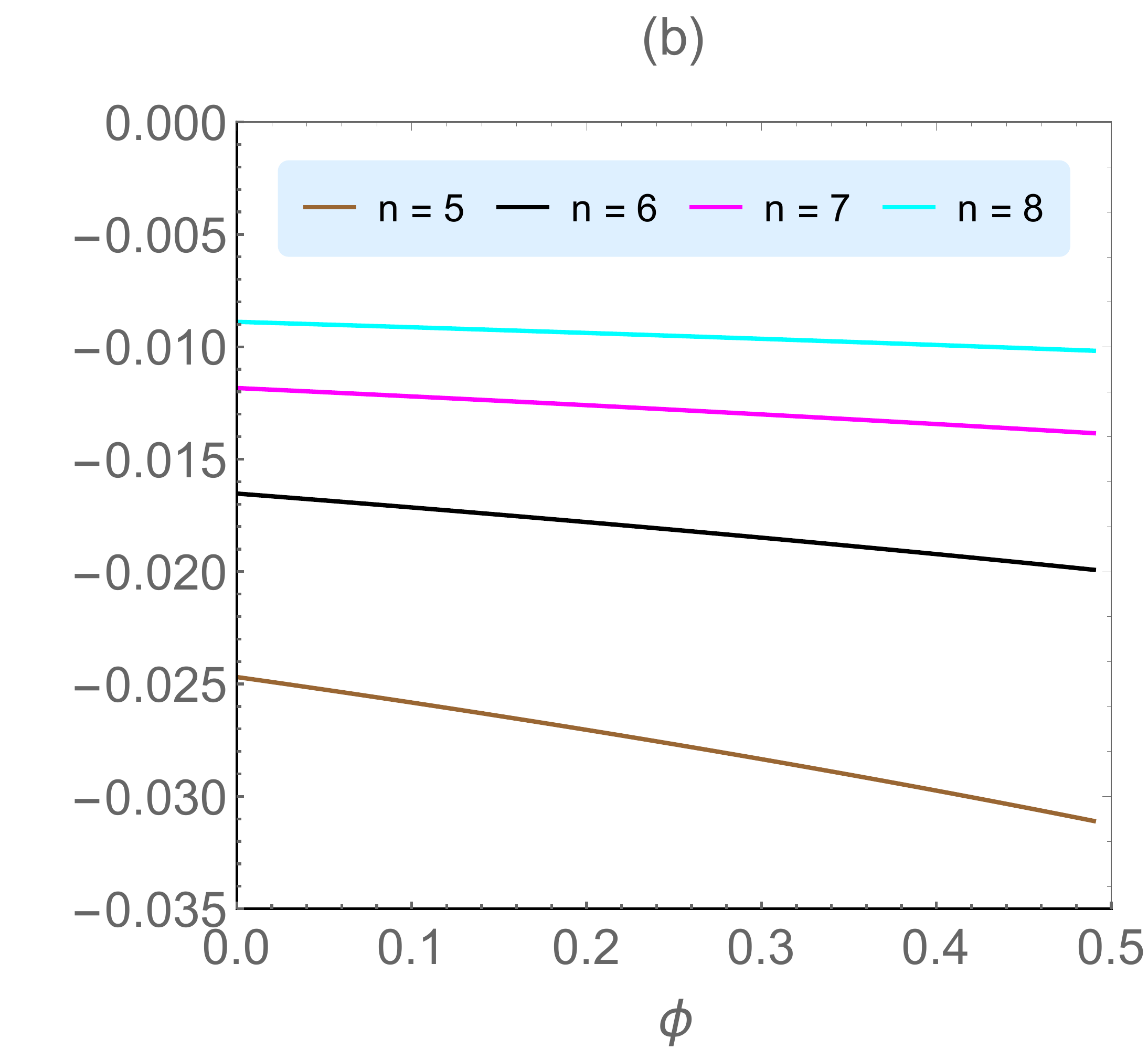}
  \caption{(Color online)
    Plots of $\mathcal{E}_{\rm irreg}$ (Eq. (\ref{Eirreg})) as a
    function of $\phi$ without taking into account rotation effects
    ($\Omega=0$).
    In (a) the first four energy levels of (1) as a function of $\phi$
    for $m=0$.
    The magnitude of the energy of the ground state is much larger than
    that of the other levels ($n=2$ (solid red line), $n=3$ (solid green
    line) and $n=4$ (solid blue line)) for any value of $\phi$ in the
    range considered.
    At $\phi=0.49$, $\mathcal{E}_{irreg,\,n=1}$ (solid orange line)
    exhibits a magnitude of $-5\times10^{3}$.
    In (b), the energy levels with $n=5$ (solid brown line), $n=6$
    (solid black line), $n=7$ (solid magenta line) and $n=8$ (solid cyan
    line).
    The spacing between the energy levels with $n>5$ decreases with
    increasing $n$, and $|\mathcal{E}_{\rm irreg}|$ increases with $\phi$.}
  \label{fig:fig3}
\end{figure*}

\begin{figure*}[!t!]
  \centering
  \includegraphics[scale=0.3]{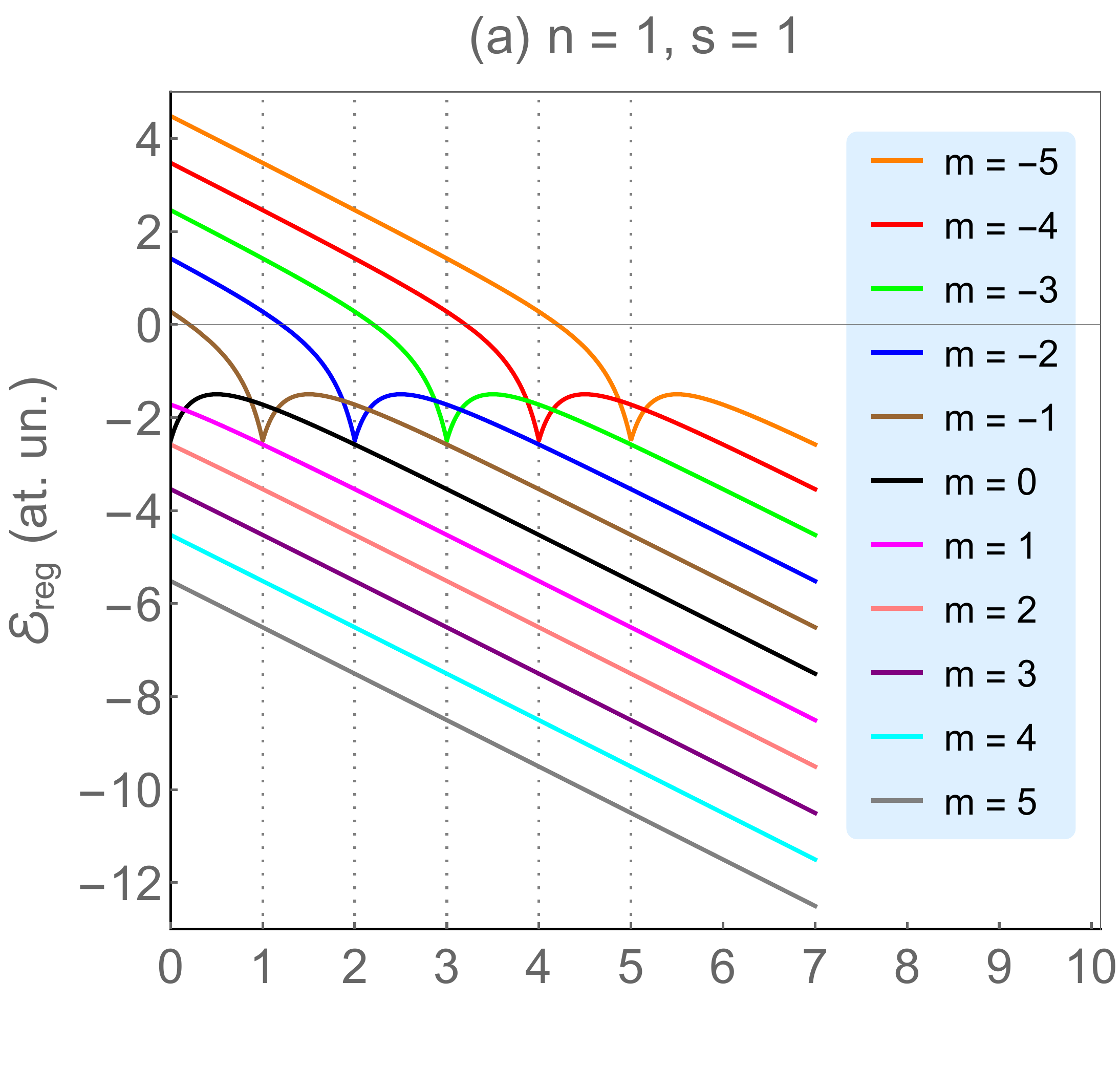}\hspace{1cm}
  \includegraphics[scale=0.3]{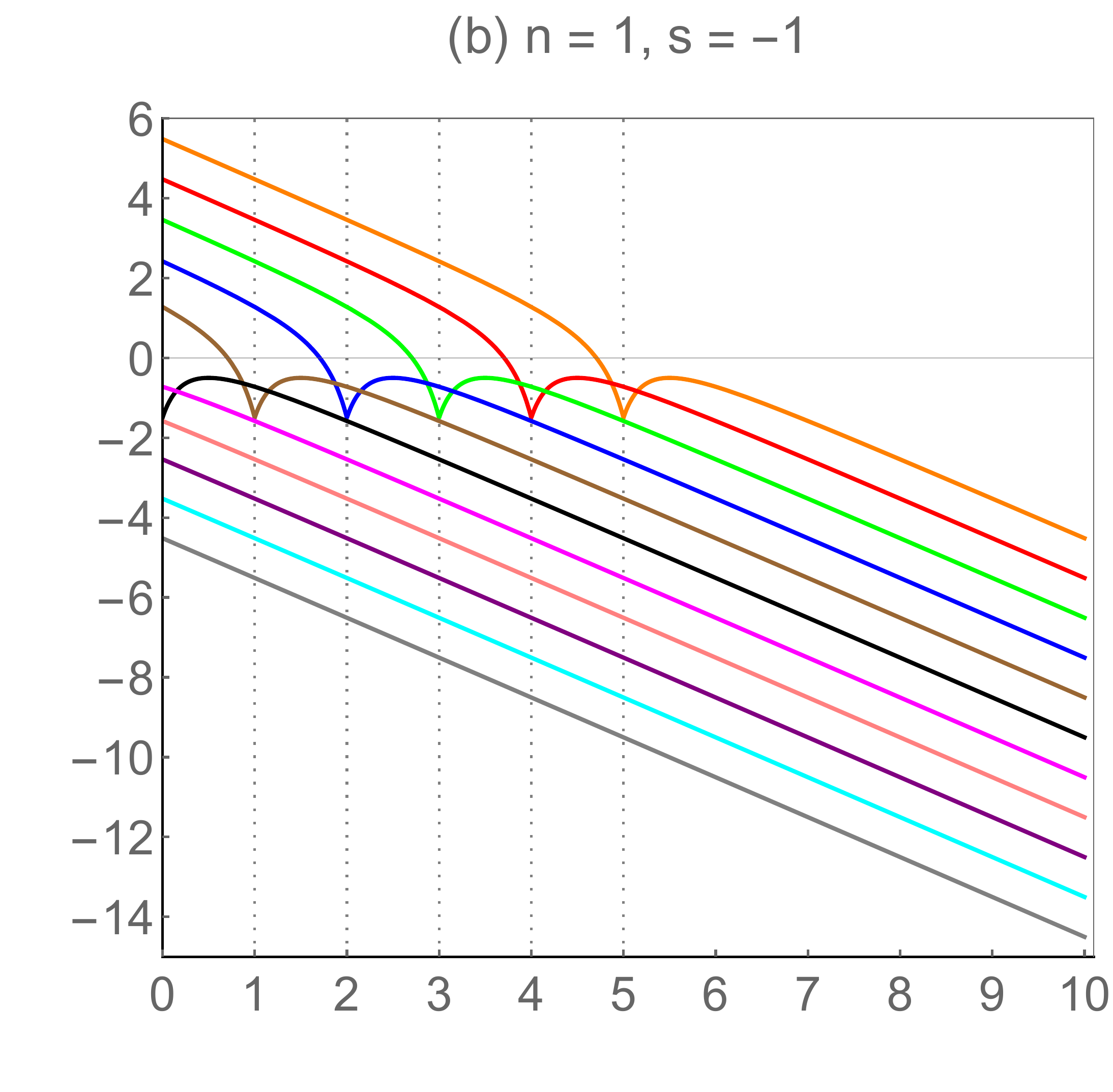}
  \includegraphics[scale=0.3]{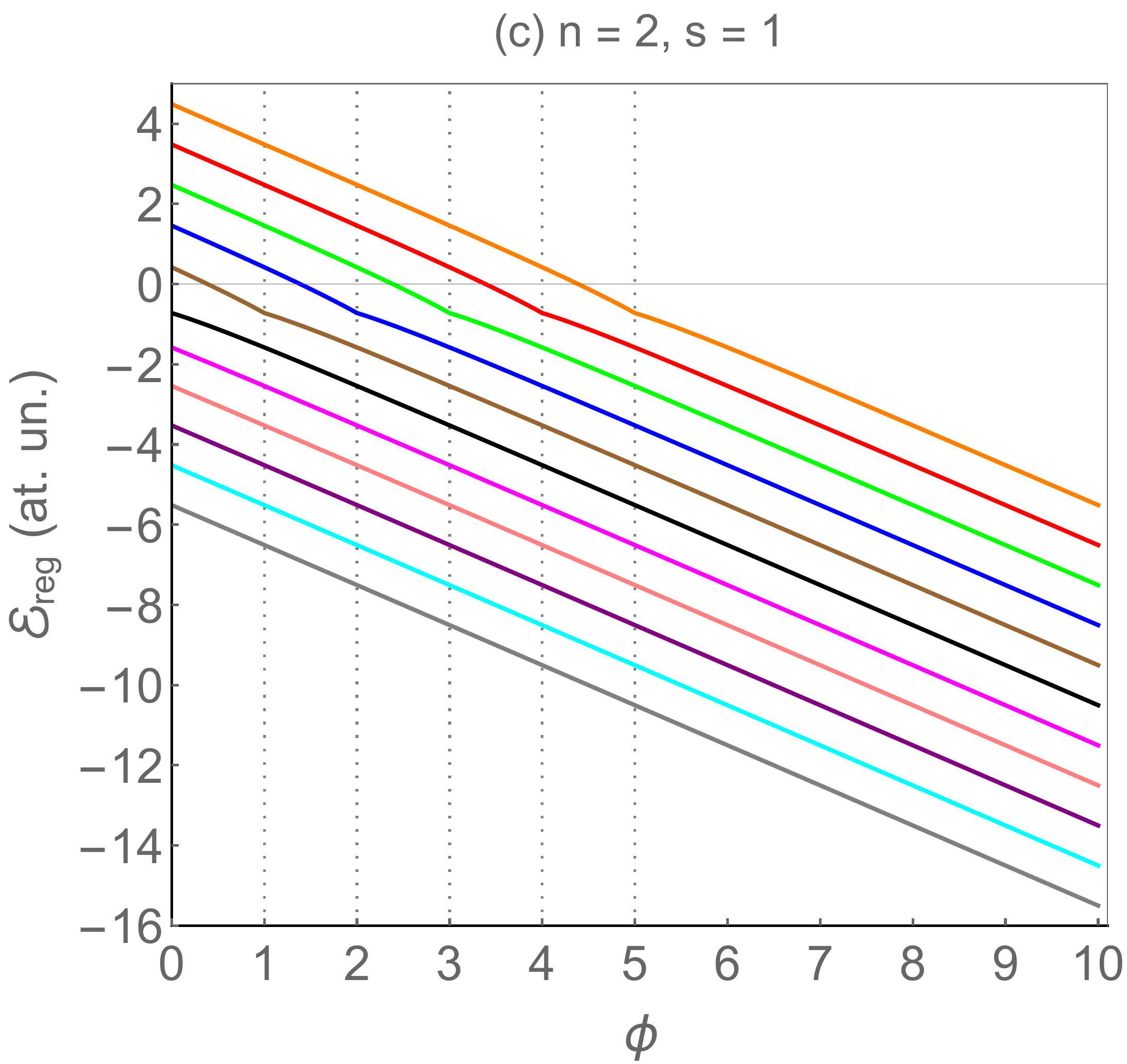}\hspace{1cm}
  \includegraphics[scale=0.3]{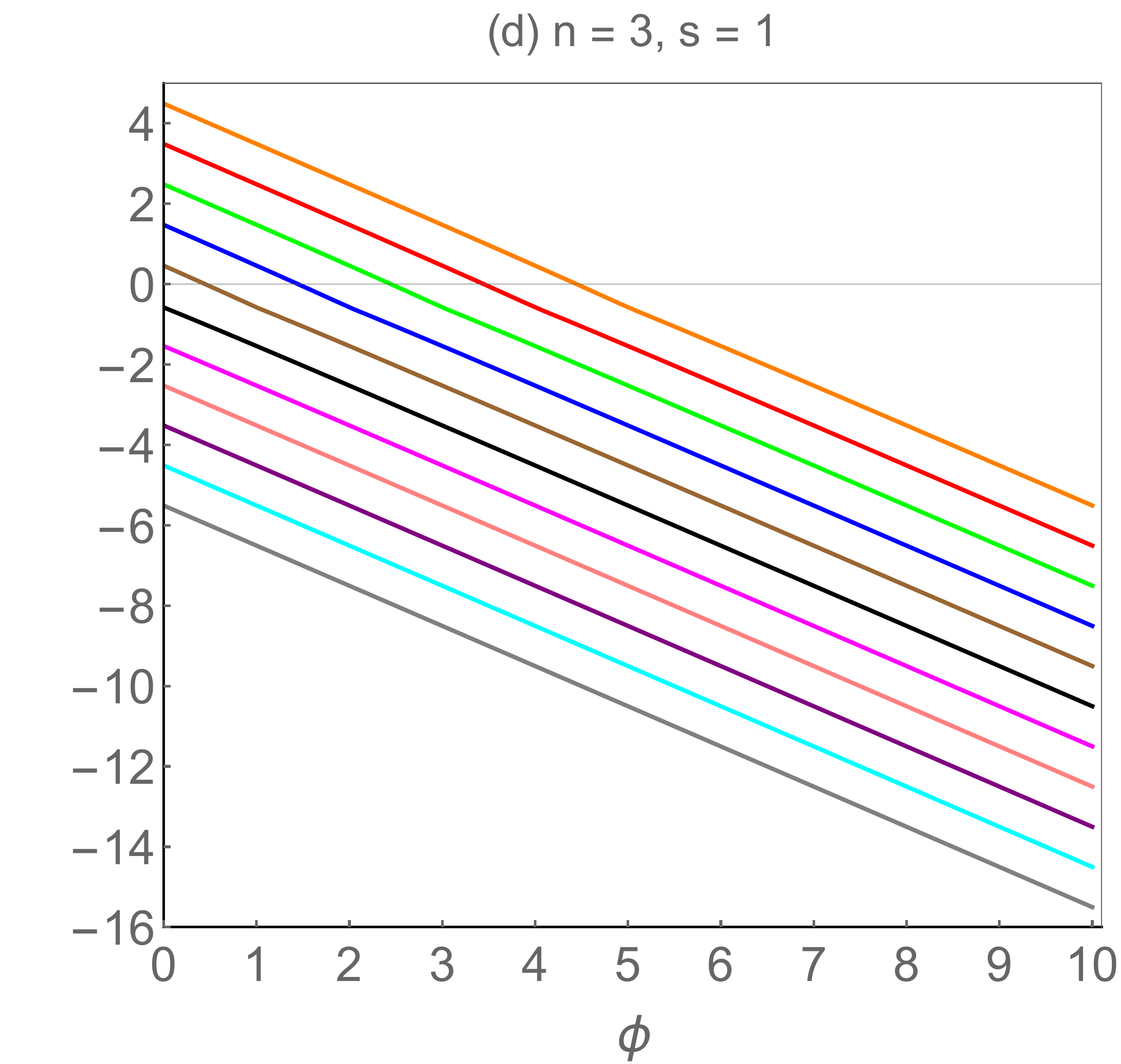}
  \caption{(Color online)
    Plots of $\mathcal{E}_{\rm reg}$ (Eq. (\ref{Ereg})) as a function of
    $\phi$ considering effects of rotation for some values of $m$.
    The rotation has the effect of shifting the energy spectrum,
    allowing positive energies values.
    The energy states with $n=1$ and $m\leqslant0$ exhibit a partially
    linear profile due to the combined effects between the magnitudes of
    the Coulomb-type potential and rotation.
    In (a), the profile for $s=+1$ and (b) for $s=-1$.
    In the first excited state ($n=2$), the effects due to the
    Coulomb-type potential are weak (panel (c)). For $n\geqslant 3$, the
    rotation effects are predominant in all range of $\phi$, and the
    profile is completely linear (panel (d)).
    Independent of the values of the parameters, we see that all energy
    levels with $m>0$ are negative and $\mathcal{E}_{\rm reg}$ increases
    with $\phi$ while for states with $m<0$, $\mathcal{E}_{\rm reg}$
    decreases with $\phi$.
  }  \label{fig:fig4}
\end{figure*}
Finally, combining the relations (\ref{ns}) and (\ref{rl2}), we obtain
\begin{equation}
\lambda _{m}\left( 2\kappa \right) ^{| j| }\frac{\Gamma
\left( 1-2| j| \right) }{\Gamma \left( \frac{1}{2}
-l_{-}\right) }+\left( 2\kappa \right) ^{-| j| }\frac{
\Gamma \left( 1+2| j| \right) }{\Gamma \left( \frac{1}{2}
+l_{+}\right)}=0.
 \label{slf}
\end{equation}

For a given value of the self-adjoint extension parameter $\lambda_{m}$ it is possible to determine an expression for the energy of the particle from the poles of the functions
$\Gamma \left(1/2-l_{-}\right)$ and $\Gamma \left(1/2+l_{+}\right)$ \cite{AoP.2010.325.2529,EPJC.2014.74.3187}.
The values of $\lambda_{m}$ we have chosen are $\lambda_{m}=0$ and $\lambda_{m}= \infty$, which are justified by the requirement that the wave function must be well behaved at these limit values.
For the choice $\lambda_{m}=0$, it means that the Hamiltonian is free of singularities.
In this case, the regular solution is the bound state wave function.
On the other hand, if $\lambda_{m}=\infty$, only the irregular solution contributes to the bound state wave function.

\section{Discussion of the results}
\label{sec:discussion}

For all other values of $\lambda
_{m}$, both regular and irregular solutions contribute to the bound state wave function.
The expressions for the energies corresponding to each case are given by
\begin{align}
  \mathcal{E}_{\rm reg} = {}
  &
    -\frac{1}{2\hbar ^{2}}\frac{m_{e}\eta ^{2}}{\left( n-
\frac{1}{2}+| m+N + \beta | \right) ^{2}}\notag\\&-\hbar \Omega \left(m+N + \beta +\frac{s}{2}\right),\;\;(\text{for }\lambda _{m}=0),\label{Ereg} \\
  \mathcal{E}_{\rm irreg} = {}
  &
    -\frac{1}{2\hbar^{2}}\frac{m_{e}\eta^{2}}{\left(n-
\frac{1}{2}-| m+N + \beta | \right)^{2}}\notag \\&-\hbar \Omega \left(m+N + \beta +\frac{s}{2}\right),\;\;(\text{for}\;\lambda_{m}=\infty),\label{Eirreg}
\end{align}
with $n=1,2,\ldots$, $N=0,1,2,\ldots $ and  0 $\leq \beta < 1$.
Notice that the rotation implies a shift in the energy levels.
 Depending on the values of $m$ and $s$, this shift can be either up or down.
For $\Omega=0$, $\hbar=1$, and then readjusting the notation for the parameters, we recover the expressions for the energies obtained in Refs.
\cite{PRD.1993.48.5935,PRD.50.7715.1994,AoP.1996.251.45}.
In the absence of the spin degree of freedom, we recover the result for
the case of regular solution at the origin \cite{TEPJD.2011.62.361}.

\begin{figure*}[!t]
  \centering
  \includegraphics[scale=0.63]{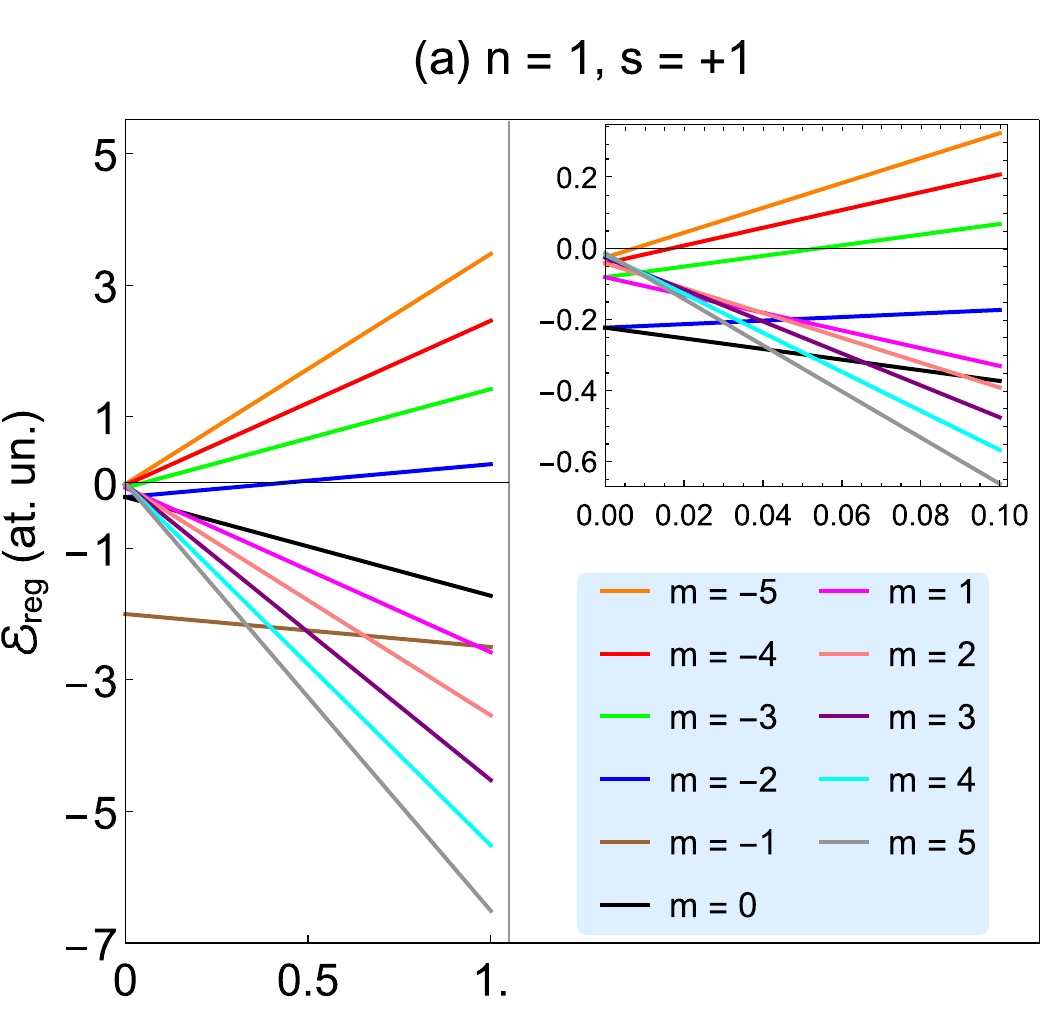}\hspace{1cm}
  \includegraphics[scale=0.63]{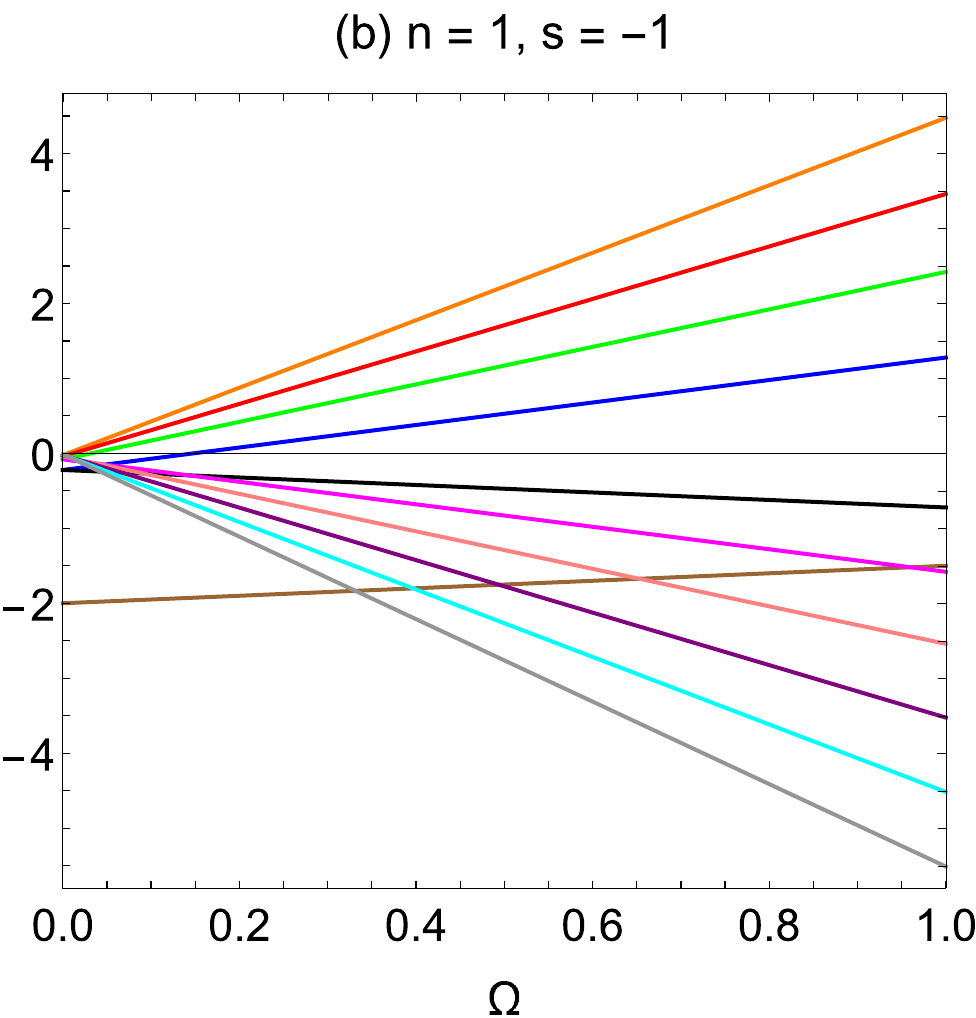}
  \includegraphics[scale=0.63]{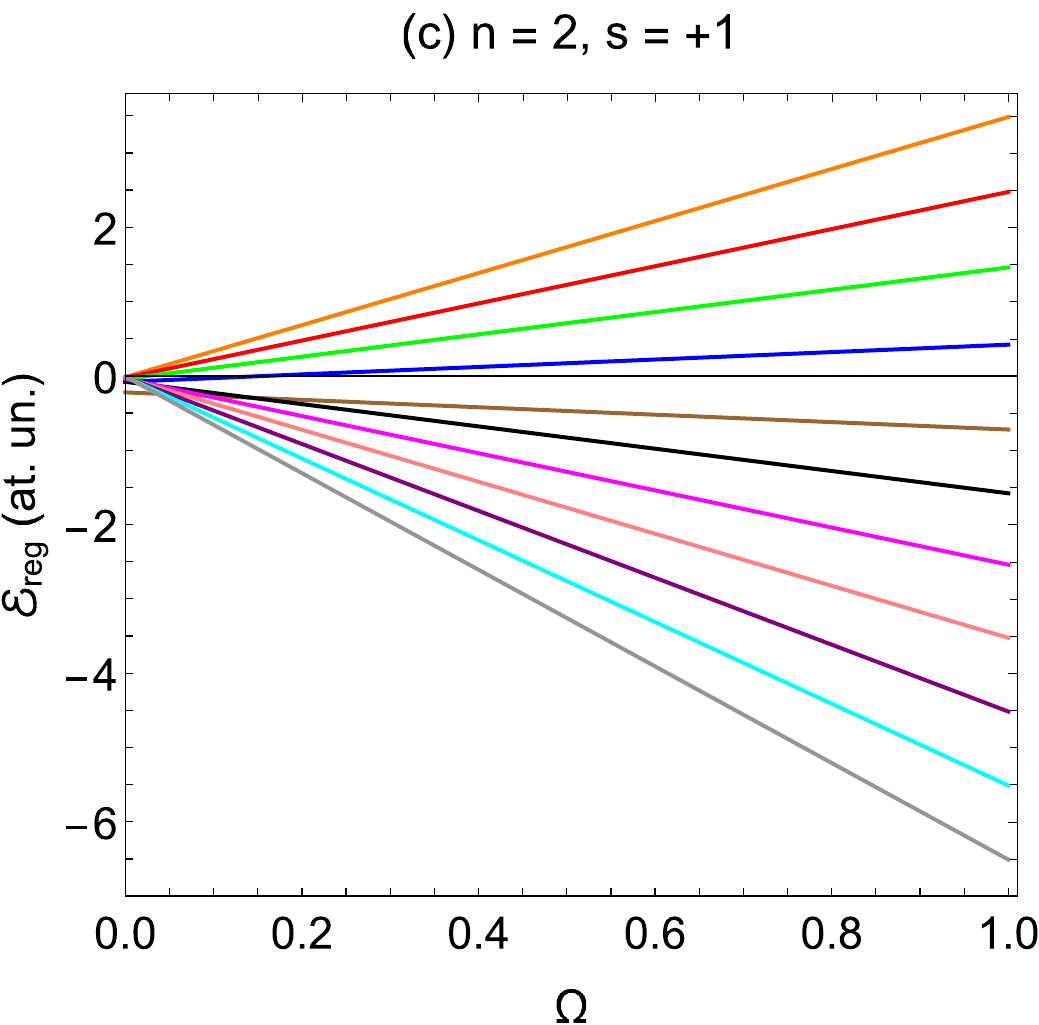}\hspace{1cm}
  \includegraphics[scale=0.63]{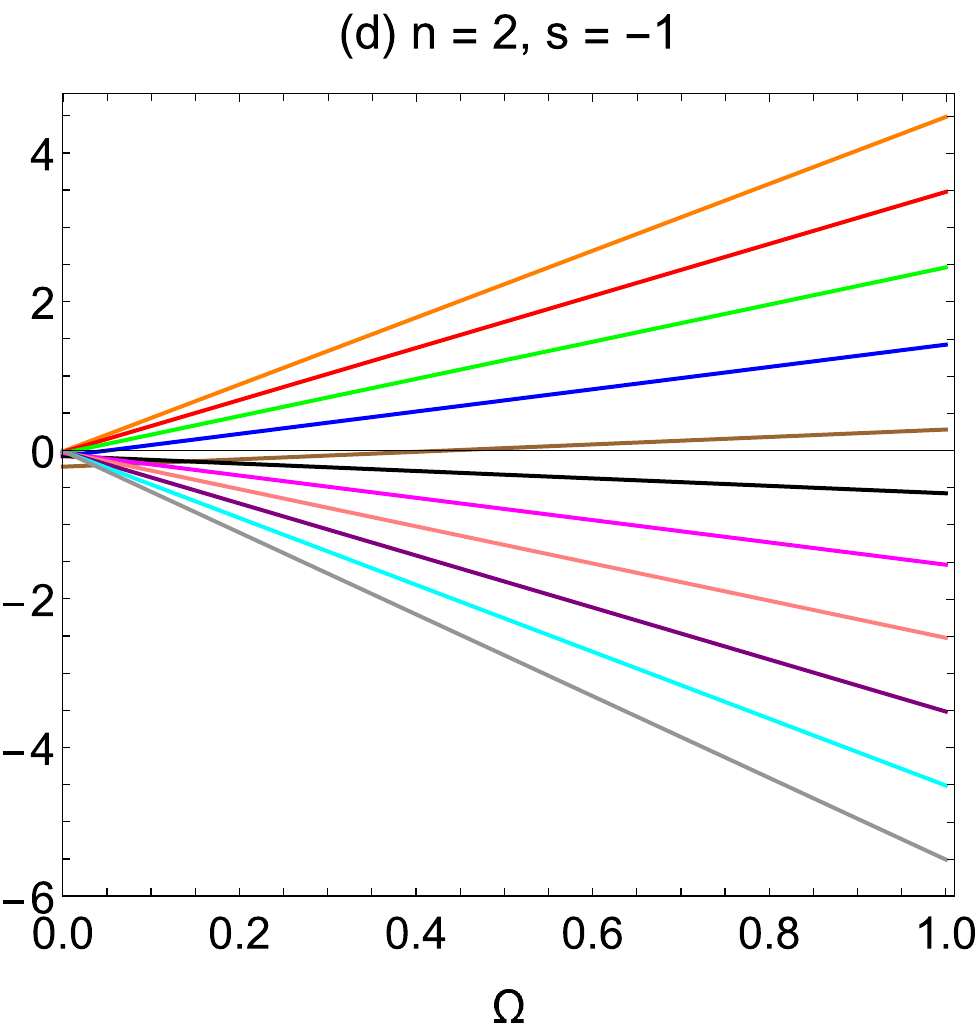}
  \caption{(Color online)
    Plots of $\mathcal{E}_{\rm reg}$ (Eq. (\ref{Ereg})) as a function of $\Omega$ for $n=1$, $n=2$, $s=\pm1$ and some values of $m$.
    In all profiles, we have a linear behavior along the range of $\Omega$. All energies with $m=[-5,+5]$ have negative values for flux near
    zero (see the inset in Fig. (a)).
    The energy of the ground state with $m=-1$ (Brown solid line) and $s=+1$ has all negative values while for $s=-1$ these energy levels
    can take on both positive (the region with these values is not displayed on the plot) and negative values (panel (b)). For the first excited state (Figs. (c)  and (d)), this characteristic is still present in the spectrum, and $|\mathcal{E}_{\rm reg}|$ increases with $\Omega$, but
    $|\mathcal{E}_{\rm reg}>0|< |\mathcal{E}_{\rm reg}<0|$. In (d), we see more clearly that the energy with $m=-1$ takes on both positive
    and negative values.
  }
\label{fig:fig5}
\end{figure*}
In this section, we study in detail the effects of the spin degree of
freedom, the AB flux, and the intensity of the Coulomb potential on the
energy levels of the particle.
Analyzing the expressions for the energies (\ref{Ereg}) and
(\ref{Eirreg}), we see that they depend explicitly on the quantities
$n$, $m$, $\phi$, $\eta$, $\Omega$ and $s$.

\begin{figure*}[!t]
  \centering
  \includegraphics[scale=0.24]{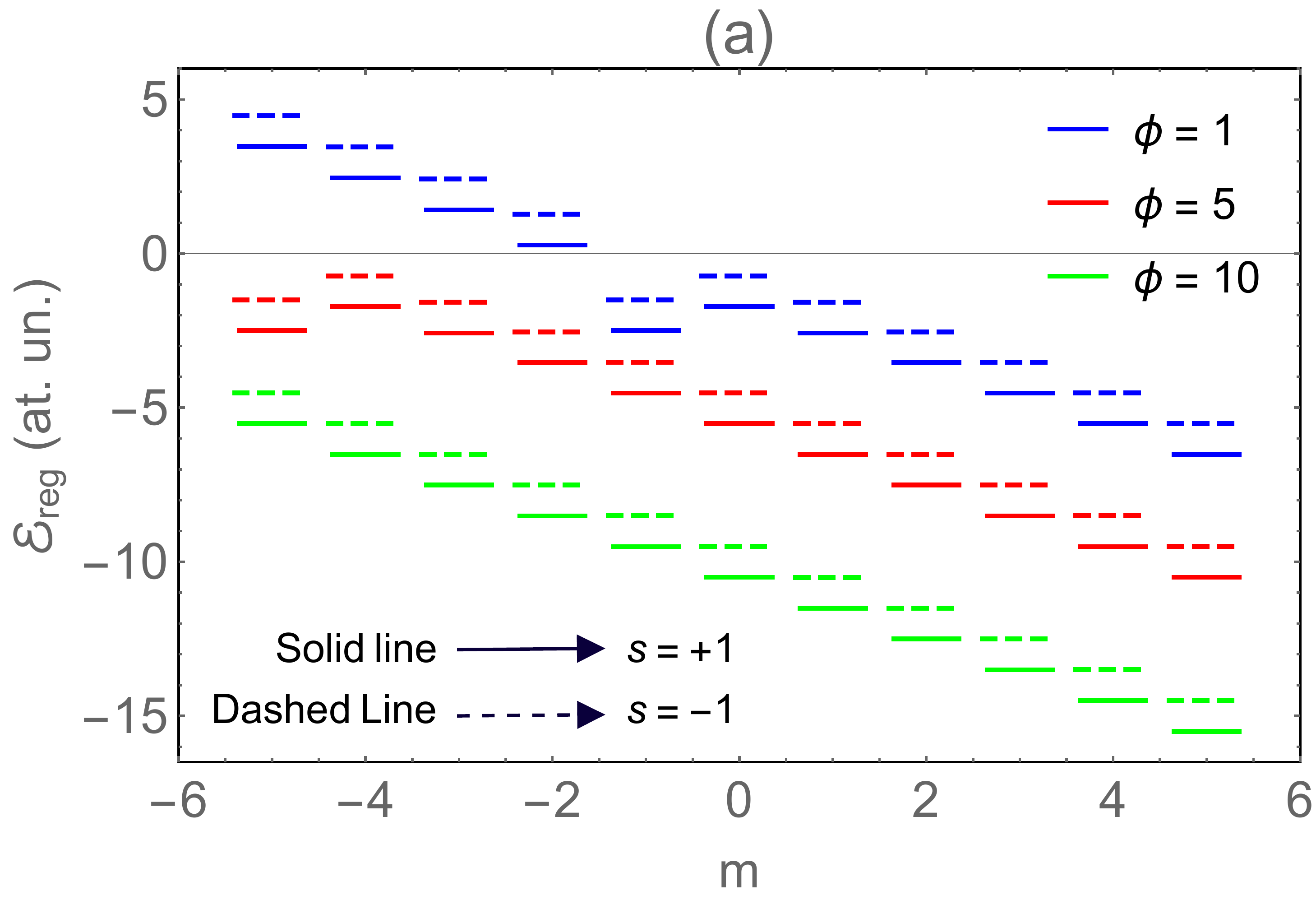}\hspace{1cm}
  \includegraphics[scale=0.24]{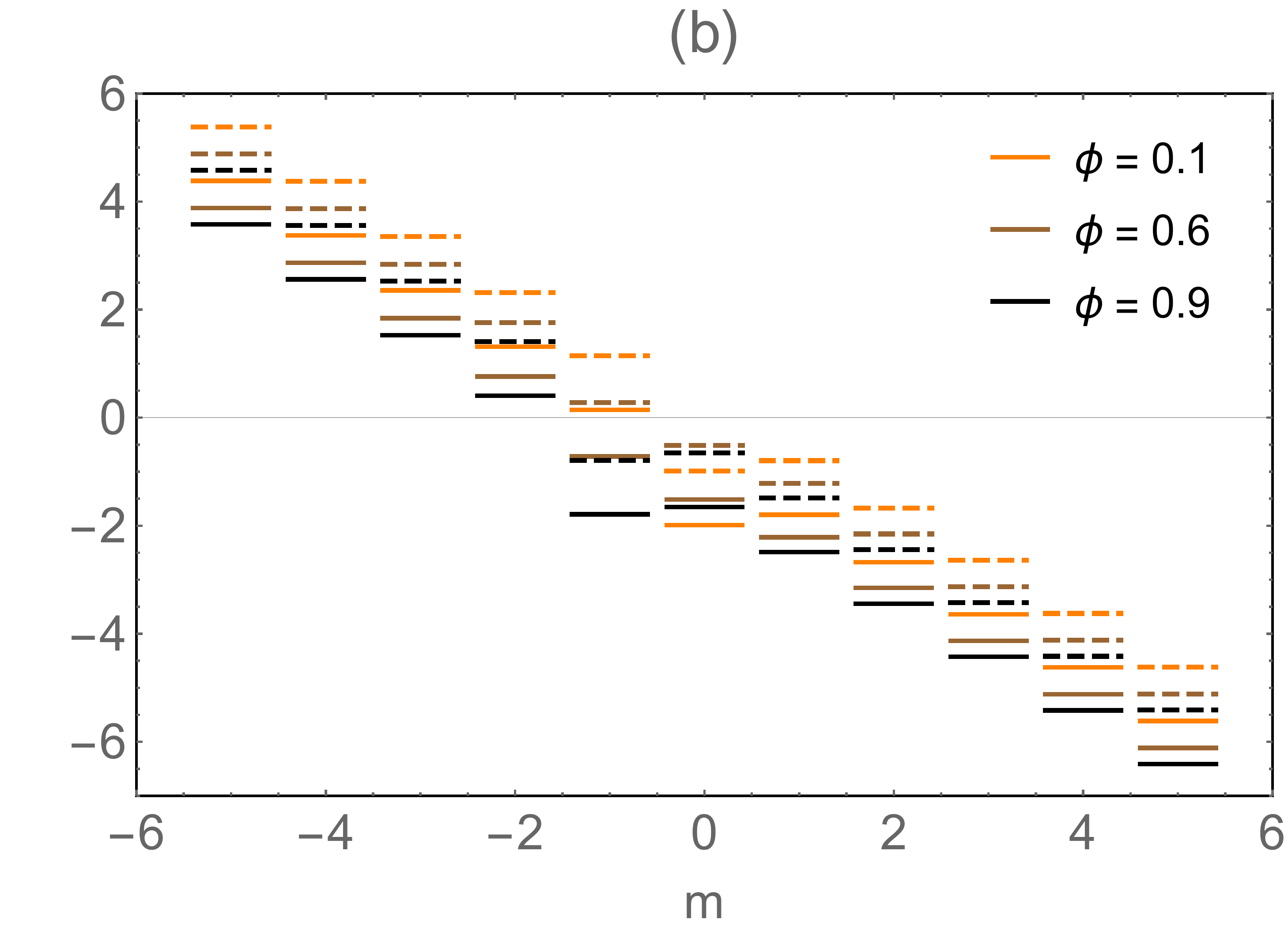}
  \caption{(Color online)
    Discrete plots of $\mathcal{E}_{\rm reg}$ (Eq. (\ref{Ereg})) as a
    function of $m$ for some values of $\phi$.
    In Fig. (a), we use $\phi=1$ (blue color), $\phi=5$ (red color) and
    $\phi=10$ while in Fig. (b), we assume $\phi=0.1$ (orange color),
    $\phi=0.6$ (brown line) and $\phi=0.9$ (black color). 
    In both profiles, the solid lines denote the states with $s=+1$ and
    the dashed lines the states with $s=-1$. 
    Degenerate states are present when we compare the profiles for the
    energies different values of element of spin along the horizontal
    axis ($m$ axis).
  } 
  \label{fig:fig6}
\end{figure*}

\begin{figure*}[!t]
  \centering
  \includegraphics[scale=0.3]{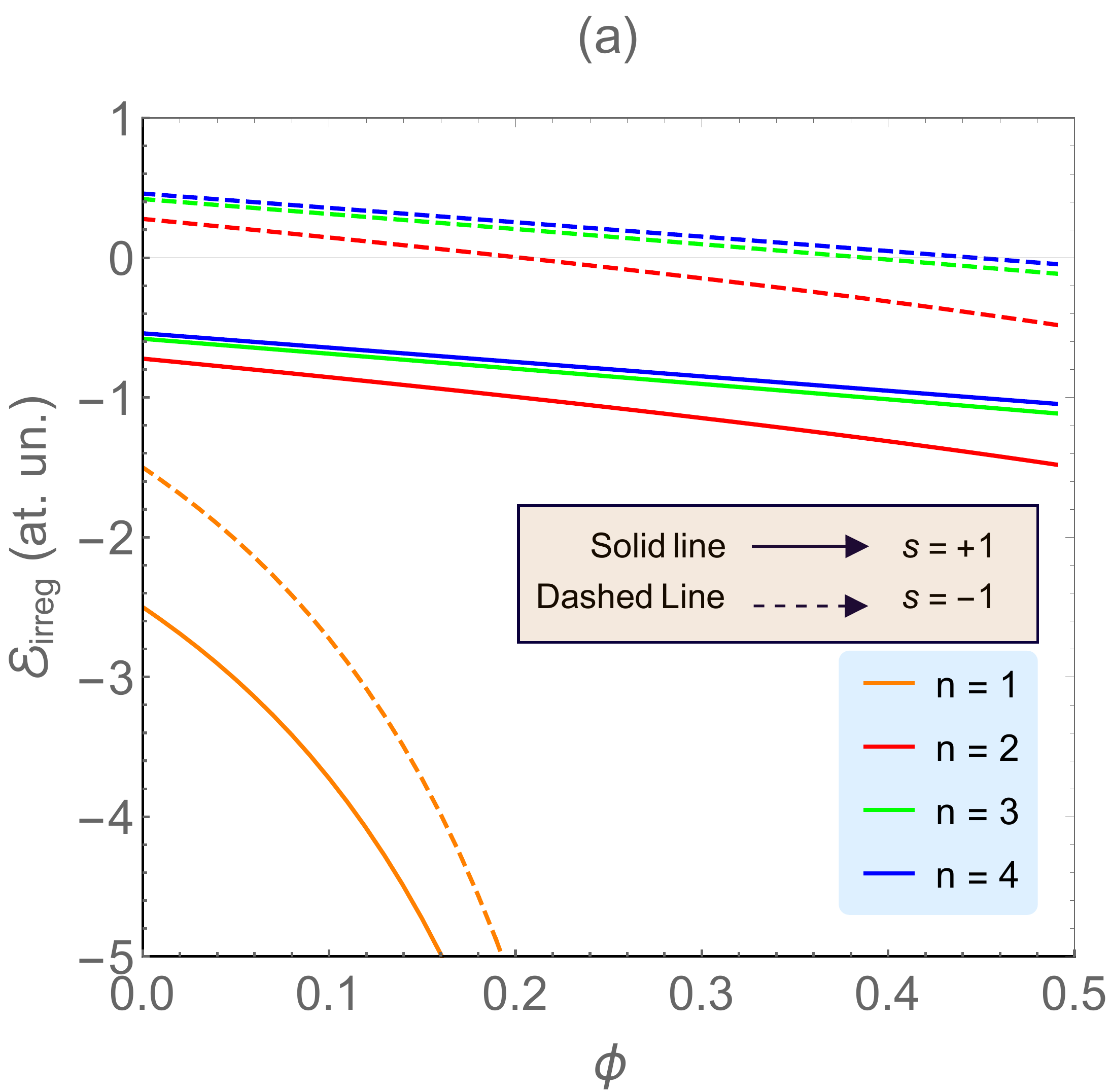}
  \includegraphics[scale=0.3]{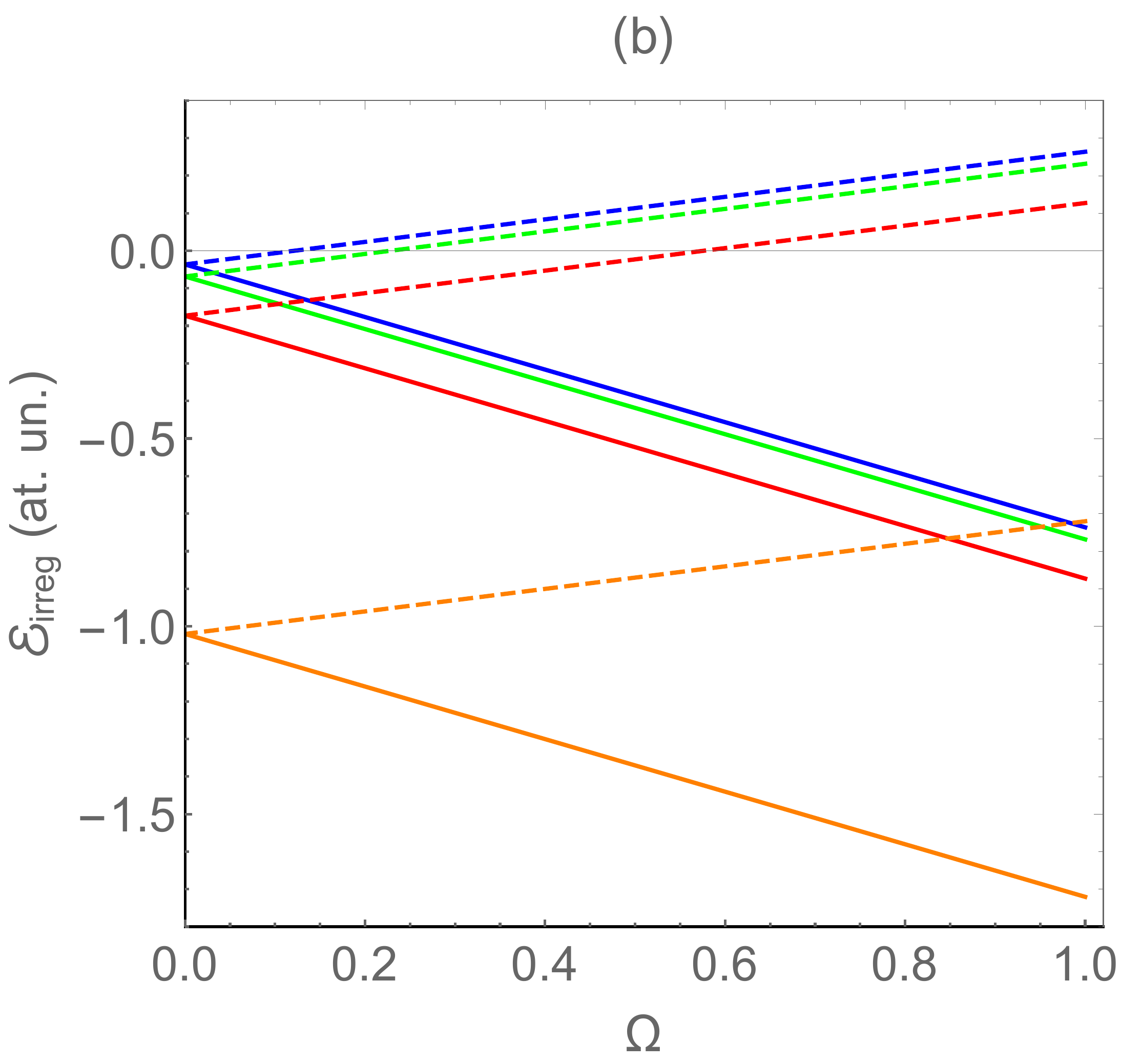}
  \caption{(Color online)
    Plots of $\mathcal{E}_{\rm irreg}$ (Eq.(\ref{Eirreg})) as a function of
    $\phi$ for $m=0$ (Fig. (a)) and as a function of $\Omega$ (Fig. (b))
    for $s=\pm 1$ and some values of $n$.
    In (a), the ground state for $s=\pm1$ (solid and dashed orange lines)
    increases to the value $\phi=0.49$.
    The energies (for $s=+1$) with $n=2$ (solid red line), $n=3$ (solid
    green line) and $n=4$ (solid blue line) are negative and have
    predominantly linear behavior with $\phi$.
    In Fig. (b), the profile of $\mathcal{E}_{\rm irreg}$ is linear with
    $\Omega$.
    For $s=+1$, all energies are negative.
    For $s=-1$, only the energy with $n=1$ is fully negative, while the
    other energy levels contain both positive and negative values.}
  \label{fig:fig7}
\end{figure*}
We shall show in detail, through graphical illustrations, that the system can assume different configurations of states of energies for a specific set of these parameters.
An immediate realization that can be verified in (\ref{Ereg}) and
(\ref{Eirreg}) is that in the absence of rotation, the resulting
equations are just the results known in the literature, from which is
found that all energy eigenvalues are negative.
Under rotation effects, we see that the energies contain an additional
linear term that explicitly depends on $\Omega$.
With the presence of this term, all energy levels are shifted, either up
or down, depending on the parameter values.
An interesting feature in this model that we will show later is that
this shift leads to some positive energy values.
Without a more detailed analysis, this feature is not so evident when we
look at the expressions for the energies.
In all our figures and discussions below, we use atomic units $\hbar=1$,
$\eta=1$ and $m_{e}=1$.

We first investigate the energy as a function of the AB flux $\phi$ for
some values of $m$ and $n$ (see Fig. \ref{fig:fig1}) for $\Omega=0$.
For $n=1$ and $m\geqslant0$, we see that $|\mathcal{E}_{\rm reg}|$
increases for smaller values of $m$, which allows us to confirm that the
state of the system of highest energy is the one with $m=0$.
For $n=2$, the curves maintain the same behavior, implying only a
decreasing in $|\mathcal{E}_{\rm reg}|$.
For values of $\phi$ close to $10$, the energies tend to zero.
For $n=1$ and $m \leqslant 0$, we observe a periodic pattern in the
energies,  which occurs whenever the flux takes on an integer value and
the minimum allowed energy is $\mathcal{E}_{\rm reg}=-2.0$
[see Fig. \ref{fig:fig2}(a)].
We also see the presence of degenerate states for different values of
$m$, and their number increases when we consider more states and the
degenerated energies are schematically indicated by black circles.
For $n=2$ and $m \leqslant 0$, the pattern of the energy curves is
similar but presents a reduction in the scale.
It implies a change in the location of the degeneracy points
[see Fig. \ref{fig:fig2}(b)].
We observe no change in the number of degenerate states in this case.
The position of the degenerate states suffers a shift because of the
change in energy scale, from the state $n=1$ to the state $n=2$.

To analyze the plot of $\mathcal{E}_{\rm irreg}$ as a function of $\phi$
for $\Omega=0$, we must take into account the allowed range for
$|j|=|m+\phi|$ in Eq. (\ref{rj}).
It leads to a reduction in the range of $\phi$ to values smaller than
$0.5$ (see Fig. \ref{fig:fig3}).
Consequently, the only value allowed for $m$ is $0$.
In Fig. \ref{fig:fig3}(a), we plot the first four energy levels of
$\mathcal{E}_{\rm irreg}$ as a function of $\phi$.
Notably, one can see that only the ground state (solid orange line) has
a magnitude that increases rapidly with $\phi$, following this pattern
up to the value $\phi=0.49$.
For energy levels with $n=2$ (red solid line), $n=3$ (green solid line)
and $n=4$ (blue solid line), we see that they have the same order of
magnitude.
The inset in the figure further illustrates this feature.
For energy levels with $n\geqslant 5$, we observe a tendency for them to
approach each other, with $|\mathcal{E}_{\rm irreg}|$ increasing with $\phi$
[see Fig. \ref{fig:fig3}(b)].

When including rotation effects, i.e., for $\Omega>0$, depending on the
values of the parameters involved, both $\mathcal{E}_{\rm reg}$ and
$\mathcal{E}_{\rm irreg}$ can assume positive or negative values.
In Fig. \ref{fig:fig4}, we plot the ground state ($n=1$) for $s=\pm 1$
[see Figs. \ref{fig:fig4}(a) and \ref{fig:fig4}(b), respectively], and
the first  ($n=2$) and second ($n=3$) excited states for $s=+1$ [see
Figs. \ref{fig:fig4}(c) and \ref{fig:fig4}(d), respectively] in the
range $m=[-5,+5]$.
Since the term involving the rotation in the energies (\ref{Ereg}) and
(\ref{Eirreg}) is linear, its effect corresponds to a shift in the
energy scale, depending on the values of the parameters involved.
All energies with $m<0$ start from positive values and decrease as the
flux is increased.
This behavior continues until the flux reaches an integer value
(represented by vertical dashed lines in Fig \ref{fig:fig4}).
Continuing by increasing the flux, we observe a slight tendency of
decreasing in the $|\mathcal{E}_{\rm reg}|$,  until it returns again to
its previous behavior.
Note that this effect occurs for all energy levels with $m<0$.
This occurs most explicitly in Figs. \ref{fig:fig4}(a) and
\ref{fig:fig4}(b).
It can be clarified by analyzing the behavior of
$\left(n-\frac{1}{2}+ |m+N + \beta| \right)^{-2}$ together with the flux
decomposition $\phi = N + \beta$ defined in Eq. (\ref{fq}).
In fact, this profile was displayed in Fig. \ref{fig:fig2} where we
analyzed the case $\Omega=0$.
For states with $m>0$, $|\mathcal{E}_{\rm reg}|$ increases linearly
as a function of $\phi$.
It is also important to note that when $s=-1$ the energy levels are only
shifted [see Fig. \ref{fig:fig4}(b)].
When we access the energies of states with $n=2$, the profile of
$\mathcal{E}_{\rm reg}$ tends to be partially linear (observe the
regions where the flux assumes integer values in
Fig. \ref{fig:fig4}(c)), while for states with $n\geqslant3$ it is predominantly linear [see Fig. \ref{fig:fig4}(d)].

As argued above, the term that explicitly depends on rotation in the
energies (\ref{Ereg}) and  (\ref{Eirreg}) is linear. 
Therefore, this term should be dominant in the spectrum for certain set
of parameters.
In Figs. \ref{fig:fig5}(a)-(b) we plot the ground state and the first
excited state in Figs. \ref{fig:fig5}(c)-(d) of $\mathcal{E}_{\rm reg}$
as a function of $\Omega$ for $s=\pm1$ and some values of $m$.
As we can see, $|\mathcal{E}_{\rm reg}|$ increases with $\Omega$ at
almost all energy levels.
Moreover, they all start from negative values [see the inset in Fig.
\ref{fig:fig5}(a)], but some states follow a tendency to assume positive
values (case of the states with $m\leqslant -2$) and others to negative
values (case of the states with $m\geqslant -1$).
In particular, we find that in the first excited state for $s=+1$ [solid
brown line in Fig.\ref{fig:fig5}(c)] the effects of rotation on the
state with $m=-1$ follows the tendency to assume increasingly negative
values while for $s=-1$ [solid brown line in Fig. \ref{fig:fig5}(d)] it
starts with negative values, and at the end of the range of $\Omega$ it
takes on positive values.

We also plot in (Fig. \ref{fig:fig6} the behavior of
$\mathcal{E}_{\rm reg}$ as a function of $m$ for $\Omega=1$, $s=\pm1$
and some values of $\phi$.
In Fig. \ref{fig:fig6}(a), we use $\phi=1$ (blue color), $\phi=5$ (red
color) and $\phi=10$ while in Fig. \ref{fig:fig6}(b), we assume
$\phi=0.1$ (orange color), $\phi=0.6$ (brown color) and $\phi=0.9$
(black color).
Tin this figures, the solid lines denote the states with $s=+1$ and the
dashed lines the states with $s=-1$.
By comparing the profiles of the solid and dashed lines (on the
horizontal axis) at the different flux values, we can see the presence
of degenerate states when $s=\pm1$.

Since the effective angular momentum in the energies of the irregular
solution satisfies relation (\ref{int}), then the most interesting plots
we can realize are the ones as a function of the AB flux and the
rotation. 
In Fig. \ref{fig:fig7}(a), we plot $\mathcal{E}_{\rm irreg}$ as a
function of $\phi$ (in the range $|\phi| \leqslant 0.49$) for $m=0$ and
$s=\pm 1$.
We can see that the ground state (solid and dashed orange lines)
increases with $\phi$ up to the value $0.49$, point in which it
exhibits a magnitude of the order of $-5\times10^{3}$.
For energy levels with $n \geqslant2$ (solid red, green and blue lines)
and $s=+1$, the energies are negative and $|\mathcal{E}_{\rm irreg}|$
increases with $\phi$. 
In contrast, when $s=-1$, we find positive energies (dominance of
rotation effects), with $|\mathcal{E}_{\rm irreg}|$ decreasing until it
reaches zero magnitude (dashed red, green and blue lines). 
Continuing by increasing the flux, we find that
$|\mathcal{E}_{\rm irreg}|$ increases until the end of the range of
$\phi$.  
Finally, when we plot the profile of $\mathcal{E}_{\rm irreg}$ as a
function of $\Omega$, we see a linear behavior (Fig.\ref{fig:fig7}(b)).
To make this plot agree with relation (\ref{int}), we use $m=0$ and
$\phi=0.2$.
Note that $|\mathcal{E}_{\rm irreg}|$ for $n=1$ and $s=+1$ increases
when $\Omega$ is increased, with $|\mathcal{E}_{\rm irreg}|$ being the
most energetic state. 
On the other hand, the profile of $\mathcal{E}_{\rm irreg}$ for $s=-1$
has energy levels which include both positive and negative values for 
$n=2$ (dashed red line), $n=3$ (dashed green line) and $n=4$ (dashed
blue line) in Fig.\ref{fig:fig7}(b).
In this configuration, only the ground state (dashed orange line) has
negative energy values, and therefore $\mathcal{E}_{\rm irreg}$
decreases with $\phi$.

From the above discussions and graphical illustrations, it is evident
that both the spin degree of freedom and rotation has crucial physical
implications on its energy spectrum. 
The combined effects between centrifugal forces and spin element
projection lead to various changes in energy levels, such as the
appearance of positive energy eigenvalues and localized peaks of low and
high magnitudes of energies.

\section{Conclusions}
\label{sec:conclusions}

In this paper, we have studied the spin-1/2 Aharonov-Bohm problem in the
presence of a Coulomb type potential in a rotating frame.
One of the main goals of this study was to investigate the role of the
spin degree of freedom, Coulomb-type interaction, and rotation on the
energy levels of the particle.
For this to be accomplished, we have considered the
Pauli-Schr\"{o}dinger equation of motion to the order of $1/m_{e}$,
which contains the spin-independent energy (kinetic energy and potential
energy) and the interaction energy (magnetic dipole energy,
spin-rotation coupling, and the energy due to inertial effects).
We have argued that the radial equation of motion contains a singularity
in the $r=0$ region.
A technique based on the self-joint extension method was used to deal
with this issue.
Then, we have solved the problem for bound states and obtained
expressions for energy levels of the particle (Eqs. (\ref{Ereg}) and
(\ref{Eirreg})).
These expressions represent the energy levels obtained from the regular
and irregular solutions at the origin, respectively.
We have investigated these energy levels as a function of the parameters
involved from several aspects.
We first analyzed the case without rotation effects, focusing only on
the effects due to the Zeeman-type term, which leads to the point
interaction at the origin. 
For the energy of the regular solution, we found that
$|\mathcal{E}_{\rm  reg}|$ for $n=1$ and $m \geqslant 0$ decreases as
the flux is increased. 
This effect occurs most quickly when the particle is in the first
excited state.
For states with $n=1$ and $m<0$, we have verified that
$\mathcal{E}_{\rm reg}\leqslant -2.0$ with the occurrence of
degenerate states while for states with $n=2$
and $m<0$, we found that $\mathcal{E}_{\rm reg}\leqslant -0.25$ and with
the same number of degenerate states.
For the energy of the irregular solution $\mathcal{E}_{\rm irreg}$, we
have shown that the states exhibiting the largest magnitudes are the
states with $n\leqslant 4$.
When we take into account the effects of rotation, $\mathcal{E}_{\rm
  reg}$ as a function of $\phi$ exhibits both positive and negative
eigenvalues for $m\leqslant-1$ and only negative eigenvalues for
$m\geqslant 0$.
We also argued that only the ground state contains degenerate states.
We have also analyzed the profile of $\mathcal{E}_{\rm reg}$ as a
function of the rotation parameter $\Omega$ for $n=1$, $s=\pm1$ and some
values of $m$. 
We have found that for $m \leqslant -2$ both positive and negative
eigenvalues are allowed while for $m\geqslant
-1$ only negative eigenvalues appear.
These features are also present in the energies of the first excited
state for $s=1$ and $s=-1$.
We also found the presence of degenerate states in the profile of
$\mathcal{E}_{\rm reg}$ as a function of $m$ for some values of $\phi$. 
We finalized the work by investigating the profiles of
$\mathcal{E}_{\rm irreg}$ as a function of $\phi$
and $\Omega$ for
$s=\pm1$ and some values of $n$. 
In the first case, we found that only the energy of the ground state
does not exhibit a linear behavior with $\phi$.
In the second profile, as already expected, the behavior is linear with
$\Omega$, and some levels contain both positive and negative energy
values. \\

\section*{Acknowledgments}
This work was partially supported by the Brazilian agencies CAPES, CNPq
and FAPEMA.
FMA also acknowledges CNPq Grants No. 434134/2018-0, and
No. 314594/2020-5.
EOS acknowledges CNPq Grant No. 307203/2019-0, and
FAPEMA Grant No. 01852/14.
This study was financed in part by the
Coordena\c{c}\~{a}o de Aperfei\c{c}oamento de Pessoal de N\'{\i}vel
Superior - Brasil (CAPES) - Finance Code 001.

\section*{References}
\bibliographystyle{apsrev4-2}
%

\end{document}